\documentclass[aps,prb,showpacs,floatfix,superscriptaddress,twocolumn]{revtex4}

\usepackage{amsmath,amssymb,graphicx}
\usepackage{xcolor}
\usepackage[colorlinks=true,urlcolor=blue,linkcolor=blue,citecolor=blue,bookmarks=false]{hyperref}

\usepackage[final]{changes}
\setdeletedmarkup{}

\definechangesauthor[name={Christoph}, color=blue]{CR}
\definechangesauthor[name={Christian}, color={blue}]{CP}

\graphicspath{{figures/}}

\begin{document}
	
%\title{Low temperature properties of single-crystal ErB$_{2}$}
\title{Easy-plane ferromagnetism in single-crystal ErB$_{2}$ at low temperatures}

\author{C. Resch}
 \email{c.resch@tum.de}
 \affiliation{School of Natural Sciences, Technical University of Munich, D-85748 Garching, Germany}
  \affiliation{Zentrum f\"ur QuantumEngineering (ZQE), Technical University of Munich, D-85748 Garching, Germany}

\author{G. Benka}
 \affiliation{School of Natural Sciences, Technical University of Munich, D-85748 Garching, Germany}
 
\author{A. Bauer}
 \affiliation{School of Natural Sciences, Technical University of Munich, D-85748 Garching, Germany}
 \affiliation{Zentrum f\"ur QuantumEngineering (ZQE), Technical University of Munich, D-85748 Garching, Germany}
\author{E. Dilmieva}
 \affiliation{HFML-FELIX, Toernooiveld 7, 6525 ED Nijmegen, The Netherlands}
 \author{U. Zeitler}
 \affiliation{HFML-FELIX, Toernooiveld 7, 6525 ED Nijmegen, The Netherlands}
 \affiliation{Institute for Molecules and Materials, Radboud University, Heyendaalseweg 135, 6525 AJ, Nijmegen, The Netherlands.}
\author{C. Pfleiderer}
 \affiliation{School of Natural Sciences, Technical University of Munich, D-85748 Garching, Germany}
 \affiliation{Zentrum f\"ur QuantumEngineering (ZQE), Technical University of Munich, D-85748 Garching, Germany}
\affiliation{Munich Center for Quantum Science and Technology (MCQST), Technical University of Munich, D-85748 Garching, Germany}  
\affiliation{Heinz Maier-Leibnitz Zentrum (MLZ), Technical University of Munich, D-85748 Garching, Germany}

\date{\today}

\begin{abstract}	
\replaced[id=CP]{We report a study of high-quality single crystals of the hexagonal rare-earth diboride ErB$_{2}$ prepared by means of the self-adjusted flux travelling-solvent floating-zone technique. Comprehensive measurements of the ac susceptibility, magnetization, specific heat, and electrical resistivity consistently establish a paramagnetic state at room temperature which develops strong easy-plane magnetic anisotropy of the hexagonal basal plane below a temperature $T_{\rm x} \approx 50$~K. Below a pronounced second-order phase transition at $T_{c} = 14$~K long-range magnetic order stabilizes. Excellent agreement with both the qualitative phenomenology and the quantitative characteristics of the thermodynamic and transport properties under applied magnetic fields up to 30\,T identifies the magnetic order in ErB$_2$ as a showcase par excellence of pronounced easy-plane ferromagnetism of Er$^{3+}$ local moments, in the presence of crystal electric fields and Van Vleck contributions.}
{We report a study of single crystals of the hexagonal rare-earth diboride ErB$_{2}$ prepared by means of the self-adjusted flux travelling-solvent floating-zone technique. Measurements of the magnetization, ac susceptibility, specific heat, and electrical resistivity consistently establish ferromagnetic order of the Er$^{3+}$ moments below a second-order phase transition at $T_{c} = 14$~K and a very strong easy-plane anisotropy. \added{The specific heat exhibits a Schottky anomaly centered around 30~K, suggesting a crystal electric field (CEF)-induced splitting of the Hunds rule ground state multiplet with an energy splitting between the ground state and the first excited CEF level of 6~meV}. Curie--Weiss fits of the ac susceptibility are characteristic of ferromagnetic coupling within the easy hexagonal basal plane, and antiferromagnetic coupling along $\langle001\rangle$. Under magnetic field within the basal plane the magnetization is reminiscent of a soft ferromagnet that is polarized in fields above a few tenth of a Tesla, vanishing hysteresis and negligible in-plane anisotropy. Under field along $\langle001\rangle$, typical hard-axis behavior is observed with the magnetization increasing only weakly up to a spin-flip transition at $\mu_{0}H_{c} = 12$~T. The easy-plane anisotropy emerges below a crossover temperature $T_{x} \approx 50$~K , i.e. a broad paramagnetic temperature range above $T_{c}$ is governed by strongly anisotropic magnetic fluctuations.}
\end{abstract}

\maketitle

%%%%%%%%%%%%%%%%%%%%%%%%%%%%%%%%%%%%%%%%%%

\section{Motivation}

\added[id=CP]{
In recent decades extensive studies of intermetallic compounds have revealed a large body of different forms of electronic order. These include, amongst others, magnetic states involving non-collinear, non-coplanar and multi-$q$ order, spin liquid and fractionalized phases under geometric frustration, polar order, or superconductivity \cite{1996_Bitko_PhysRevLett, 2007_Lohneysen_RevModPhys, 2022_Wendl_Nature, 2014_Chen_PhysRevLett, 2015_Nakatsuji_Nature, 2016_Nayak_SciAdv, 1990_Kosevich_PhysRep, 2012_Artyukhin_NatMater}. In comparison, particularly simple forms of magnetic order, such as plain ferromagnetism or Neel antiferromagnetism, are exceptionally rare. For instance, amongst the elements, superconductivity and complex forms of antiferromagnetism are abundant, whereas only four elements (Fe, Ni, Co, Gd) display room temperature ferromagnetism. Yet, these putatively simplest forms of order, especially ferromagnetism, represent a cornerstone of the overall understanding of electronic order, providing key information for the notions of symmetry breaking,  dynamics and criticality, defects and textures and the underlying interactions.}

\deleted[id=CP]{The C32 diborides, $M$B$_{2}$, crystallize in a bilayer hexagonal structure, space group $P6/mmm$. Depending on the element $M$, a wide range of electronic and magnetic properties may be observed,\cite{1963_Silver_JChemPhys, 1969_Barnes_PhysLettA, 1970_Kasaya_JPhysSocJpn, 1976_Tanaka_JLessCommonMet, 1977_Funahashi_SolidStateCommun, 2001_Gasparov_JETPLett, 2004_Takeya_PhysicaC, 2005_Nunes_ActaMater, 2007_Singh_PhysRevB, 2007_Suh_PhysRevB, 2010_Okamoto_ActaMater, 2014_Bauer_PhysRevB} with superconductivity in MgB$_{2}$ representing the perhaps most prominent example~\cite{2001_Nagamatsu_Nature, 2001_Kortus_PhysRevLett, 2002_Choi_Nature, 2002_Mazin_PhysRevLett}. In diborides based on rare-earth elements, at low temperatures typically magnetic order emerges with localized f-electron moments in the presence of strong magnetocrystalline anisotropy and geometric frustration~\cite{1977_Buschow_Book, 1979_Will_JLessCommonMet, 2003_Avila_JAlloyCompd, 2007_Novikov_JThermAnalCalorim, 2009_Mori_PhysRevB}.} 

\deleted[id=CP]{In this context, scenarios such as quantum criticality under transversal field tuning~\cite{1996_Bitko_PhysRevLett, 2007_Lohneysen_RevModPhys, 2022_Wendl_Nature}, large anomalous Hall effects in noncollinear magnets~\cite{2014_Chen_PhysRevLett, 2015_Nakatsuji_Nature, 2016_Nayak_SciAdv}, or solitonic excitations in Ising-like systems~\cite{1990_Kosevich_PhysRep, 2012_Artyukhin_NatMater} may be realized in a relatively simple crystallographic environment.}

\deleted[id=CP]{However, due to high melting temperatures, high vapor pressures, and peritectic metallurgical reactions, the preparation of rare-earth diborides can be challenging and studies of single-crystals are scarce.}

\added[id=CP]{In the development of a well-founded and deep understanding of the wide range of electronic order,  isostructural series of intermetallic compounds are of central interest, permitting to single out and identify specific mechanisms. They are of additional interest as the basis of thin-film heterostructures and tailored systems, in which members of isostructural series with different electronic properties are combined. An important example in this context is the class of C32 diborides, $M$B$_{2}$, which crystallize in a bilayer hexagonal structure, space group $P6/mmm$. Depending on the element $M$, a wide range of electronic and magnetic properties may be observed,\cite{1963_Silver_JChemPhys, 1969_Barnes_PhysLettA, 1970_Kasaya_JPhysSocJpn, 1976_Tanaka_JLessCommonMet, 1977_Funahashi_SolidStateCommun, 2001_Gasparov_JETPLett, 2004_Takeya_PhysicaC, 2005_Nunes_ActaMater, 2007_Singh_PhysRevB, 2007_Suh_PhysRevB, 2010_Okamoto_ActaMater, 2014_Bauer_PhysRevB, Pei2023_MoB2} with superconductivity in MgB$_{2}$ representing the perhaps most prominent example~\cite{2001_Nagamatsu_Nature, 2001_Kortus_PhysRevLett, 2002_Choi_Nature, 2002_Mazin_PhysRevLett}. In diborides based on rare-earth elements, at low temperatures typically \replaced[id=CR]{ferromagnetic}{complex forms of magnetic} order emerges with localized f-electron moments and polar degrees of freedom in the presence of strong magnetocrystalline anisotropy and geometric frustration~\cite{1977_Buschow_Book, 1979_Will_JLessCommonMet, 2003_Avila_JAlloyCompd, 2007_Novikov_JThermAnalCalorim, 2009_Mori_PhysRevB}. However, due to high melting temperatures, high vapor pressures, and peritectic metallurgical reactions, the preparation of rare-earth diborides can be challenging and studies of high-quality single-crystals are scarce.}

\added[id=CP]{Placed in this general context, the identification of particularly simple and well-defined forms of electronic order such as the conventional BCS high-$T_c$ superconductivity in MgB$_2$ \cite{2001_Nagamatsu_Nature, 2001_Kortus_PhysRevLett, 2002_Choi_Nature, 2002_Mazin_PhysRevLett}, or the itinerant-electron antiferromagnetism in CrB$_2$, which was recently reported to become superconducting under high pressures up to 100\,GPa, receives special attention \cite{2014_Bauer_PhysRevB,Pei2021_CrB2, Biswas2023_CrB2, Meng2024_XB2}. From the point of view of intermetallic compounds on the whole, as well as the class of C32 systems the putative identification of textbook-like prototypical properties, notably materials exhibiting plain-vanilla long-range ferromagnetism, is therefore of great interest.}

ErB$_{2}$ is a typical representative of the class of rare-earth diborides. Early studies of polycrystalline ErB$_{2}$ reported magnetic order below $\sim$16~K that was attributed to ferromagnetic interactions due to a positive Curie--Weiss temperature~\cite{1977_Buschow_Book}. A kink in the resistivity~\cite{2015_Kargin_Book}, consistent with an anomaly in the specific heat~\cite{2011_Novikov_PhysStatusSolidiB}, was attributed to an onset of magnetic order below $\sim$14~K, while the lattice constant $c$ was reported to increase below about 100~K~\cite{2010_Novikov_PhysSolidState}. The polycrystalline samples for these studies were prepared by means of high-pressure synthesis followed by annealing and contained several percent of impurity phases~\cite{2009_Matovnikov_InorgMater}.  Studies of single crystals were so far restricted to thin platelets of $1\times1\times0.01~\mathrm{mm}^{3}$~\cite{1972_Castellano_MaterResBull}. Recent studies of polycrystalline and single-crystal ErB$_2$ suggest that magnetic order is stabilized under pressures up to 5.6 GPa.~\cite{2024_Nuzhina_,Tong_2025}. Consequently, no details of the nature of the magnetic order, the magnetic field dependence, and the magnetocrystalline anisotropy had been reported. 

We report an investigation of the low-temperature properties of large \added[id=CP]{high-quality} single crystals of ErB$_{2}$. The single-crystal ingots were prepared by means of the self-adjusted flux travelling-solvent floating-zone approach, as reported in Refs.~\onlinecite{2022_Bauer_PhysStatusSolidiB, 2021_Benka_}. \replaced[id=CP]{In our study we find consistently long-range ferromagnetic order in the magnetization, ac susceptibility, specific heat, and electrical resistivity of }{Magnetic order of the} Er$^{3+}$ moments below a second-order phase transition at $T_{c} = 14$~K\deleted{is consistently observed in the magnetization, ac susceptibility, specific heat, and electrical resistivity}. \replaced[id=CR]{Curie--Weiss fits of the ac susceptibility suggest ferromagnetic interactions within the hexagonal basal plane, whereas the negative Curie--Weiss temperature along the $\langle001\rangle$ axis may reflect either antiferromagnetic interactions or, more likely, the crystal electric field.}{Curie--Weiss fits of the ac susceptibility suggest that the moments are coupled ferromagnetically in the hexagonal basal plane and antiferromagnetically along the $\langle001\rangle$ axis.}  The magnetocrystalline anisotropy is strong and may be described by a magnetic hard $\langle001\rangle$ axis and an easy hexagonal basal plane with negligible in-plane anisotropy. The paramagnetic state exhibits strongly anisotropic magnetic fluctuations up to a crossover temperature $T_{\rm x} \approx 50$\,K well above $T_{c}$. \added{The specific heat exhibits a Schottky anomaly centered around 30\,K, indicating a CEF-induced splitting of the ground state multiplet. In a two-level Schottky model involving two doublets, the position of the Schottky maximum corresponds to an energy splitting of approximately 6\,meV.}  

For magnetic fields applied within the basal plane, the field dependence of the magnetization is reminiscent of a soft ferromagnet with vanishing hysteresis. For fields along $\langle001\rangle$, the magnetization at low temperatures shows typical hard-axis behavior with a spin-flip transition at $\mu_{0}H_{c} = 12$~T. Combining measurements of the magnetization, the ac susceptibility, and the electrical resistivity for a large number of temperatures and fields, magnetic phase diagrams are inferred for magnetic fields applied along $\langle100\rangle$, $\langle210\rangle$, and $\langle001\rangle$.%, providing crucial points of reference for future studies.

\added[id=CP]{
Taking together our comprehensive thermodynamic and transport data, we find consistently that single-crystal ErB$_2$ supports a paramagnetic state at high temperatures, which develops a strong easy-plane magnetic anisotropy at low temperatures. In this paramagnetic state pronounced easy-plane long-range ferromagnetic order stabilizes in a weakly correlated good metallic state. Quantitatively, the magnetic properties of ErB$_2$ may be fully accounted for in terms of well-defined Er$^{3+}$ local moments, where the crystal electric fields lift the 16-fold multiplet degeneracy, together with Van Vleck contributions.}

Our paper is organized as follows. In Sec.~\ref{methods} the experimental methods are described. The presentation of the experimental results starts with the temperature dependence of the ac susceptibility and the specific heat in zero magnetic field in Secs.~\ref{susceptibility} and \ref{specificheat}, respectively. This is followed by the magnetic field dependence of the magnetization, addressing also demagnetization effects, and the temperature dependence of the magnetization and the ac susceptibility at high magnetic fields in Sec.~\ref{magnetization}. In Sec.~\ref{resistivity}, the temperature dependence of the electrical resistivity is reported, followed by the magnetic phase diagrams in Sec.~\ref{phasediagram}. \added[id=CP]{An in-depth discussion is presented in Sec.\,\ref{discussion} comprising the qualitative phenomenology, the electronic structure, the magnetic properties and further aspects}. Our results are  \deleted[id=CP]{discussed briefly and} summarized in Sec.~\ref{conclusion}.

%%%%%%%%%%%%%%%%%%%%%%%%%%%%%%%%%%%%%%%%%%

\section{Experimental methods}
\label{methods}

\deleted[id=CP]{As reported in Ref.~\onlinecite{2022_Bauer_PhysStatusSolidiB}, a} \added[id=CP]{A} large single-crystal ingot of ErB$_{2}$, forming in a weakly peritectic reaction, was grown by means of the self-adjusted flux travelling-solvent floating-zone approach \added[id=CP]{\cite{2022_Bauer_PhysStatusSolidiB}}. After a thorough metallurgical characterization establishing phase-pure single-crystal material, the ingot was oriented by means of x-ray Laue diffraction and samples were cut using a wire saw. Four samples were \replaced[id=CP]{investigated }{used} in  \replaced[id=CP]{our }{the present} study: (\replaced[id=CP]{S1}{i})~a cuboid of $2.5\times1.4\times1~\mathrm{mm}^{3}$ oriented along $\langle001\rangle \times \langle100\rangle \times \langle210\rangle$, (\replaced[id=CP]{S2}{ii}) a cube with an edge length of 0.5~mm oriented \replaced[id=CP]{like S1}{the same way}, (\replaced[id=CP]{S3}{iii})~a platelet of $2\times1\times0.18~\mathrm{mm}^{3}$ oriented along $\langle001\rangle \times \langle210\rangle \times \langle100\rangle$, and (\replaced[id=CP]{S4}{iv})~a platelet of $1.2\times1\times0.16~\mathrm{mm}^{3}$ oriented along $\langle100\rangle \times \langle210\rangle \times \langle001\rangle$.

Measurements of the magnetization, the ac susceptibility, and the specific heat were carried out in a Quantum Design physical property measurement system at temperatures down to 2~K under magnetic fields up to 14~T. The magnetization was measured with the standard extraction technique of the ACMS-II option. The ac susceptibility was measured with an excitation frequency of 911~Hz at an excitation amplitude of 1~mT. The specific heat was measured using a quasiadiabatic large heat pulse technique, where typical pulses had a size of 30\% of the temperature at the start of the pulse~\cite{2013_Bauer_PhysRevLett}. The specific heat data as well as the ac susceptibility in finite fields were measured on sample S1. Due to the pronounced magnetocrystalline anisotropy of ErB$_{2}$, samples can exert considerable magnetic torque under applied magnetic field, requiring the use of the smaller sample S2 for the magnetization measurements as a function of field applied along the hard $\langle001\rangle$ axis. For the sake of consistency, the ac susceptibility data for zero field and the magnetization data for finite fields within the easy basal plane were measured on the same sample S2. 

\added[id=CR]{The field dependence of the magnetization up to 30\,T was measured on sample S2 using a vibrating sample magnetometer (VSM) installed in a Bitter magnet at the High Field Magnet Laboratory (HFML-EMFL) in Nijmegen, the Netherlands. The measurements were performed using Bitter Cell 5, which provides a maximum magnetic field of 30\,T and is equipped with a flow cryostat. Measurements were carried out at 4.2\,K. Absolute values of the magnetization were inferred by comparing to PPMS data recorded at the same temperature on sample S2. Magnetization was measured for increasing and decreasing field, and the two sweeps were averaged to mitigate effects of recording lag relative to the applied field.}

Measurements of the electrical resistivity were carried out in conventional superconducting magnet systems equipped with a variable temperature insert. For these measurements, samples S3 and S4 were contacted by means of Al$_\mathrm{99}$Si$_\mathrm{1}$ wires. The electrical resistivity was measured using low-noise impedance-matching transformers (Signal Recovery Model 1900), digital lock-in amplifiers (Stanford Research SR830), and a precision current source (Keithley 6221). Data were recorded at an excitation frequency of 22.08~Hz and an excitation current  of several mA. The magnetic field was applied perpendicular to the current direction. For magnetic fields applied along the hard magnetic axis, notably sample S4, the sample was attached using Stycast 2850 FT epoxy. For all resistivity samples studied, a low residual resistivity of a few ${\rm \mu\Omega cm}$ was observed characteristic of excellent sample quality.

If not stated otherwise, data were recorded after zero-field cooling. Measurements of the magnetization as a function of field at low temperatures were carried out in terms of hysteresis loops including \added[id=CP]{virgin} magnetic \deleted[id=CP]{virgin} curves \added[id=CP]{after zero-field cooling}. At the level of the resolution of our data, no hysteresis was observed. For clarity, the magnetization data in Figs.~\ref{figure4} and \ref{figure6} are shown as a function of decreasing field only.

\deleted[id=CP]{In magnetic materials, demagnetization effects may lead to discrepancies between applied and internal magnetic fields in both field strength and orientation.} Approximating the samples as rectangular prisms~\cite{1998_Aharoni_JApplPhys}, the demagnetization factors $N$ of the samples used in our study were $N_{\mathrm{S1}}^{001} = 0.19$, $N_{\mathrm{S1}}^{100} = 0.34$, $N_{\mathrm{S1}}^{210} = 0.47$, $N_{\mathrm{S2}} = 0.33$, $N_{\mathrm{S3}}^{100} = 0.76$, and $N_{\mathrm{S4}}^{001} = 0.75$, where the subscript refers to the sample number. The internal magnetic field values $H_{\mathrm{int}}$ were calculated from the applied field $H_{\mathrm{ext}}$ using the measured magnetization $M(H_{\mathrm{ext}})$ using $H_{\mathrm{int}} = H_{\mathrm{ext}} - NM(H_{\mathrm{ext}})$. As the magnetization changes with temperature, data as a function of temperature at constant applied field do not correspond directly to data at constant internal field. To avoid ambiguities, data are therefore shown as a function of applied field, if not stated otherwise. 

In order to combine data measured on samples with different geometry into a single magnetic phase diagram, for each data point demagnetization effects \replaced[id=CP]{were determined }{are considered separately} for the given temperature and field values, cf.\ also Ref.~\onlinecite{2016_Bauer_Book}. Furthermore, in measurements of the ac susceptibility, demagnetization effects also modify the absolute value \replaced[id=CP]{of the susceptibility by virtue of corrections of }{via} the excitation field. The internal ac susceptibility, $\chi_{\mathrm{ac}}^{\mathrm{int}}$, may be calculated from $\chi_{\mathrm{ac}}^{\mathrm{ext}}$, using $\chi_{\mathrm{ac}}^{\mathrm{int}} = \chi_{\mathrm{ac}}^{\mathrm{ext}} / (1 - N\chi_{\mathrm{ac}}^{\mathrm{ext}})$, as discussed below.

%%%%%%%%%%%%%%%%%%%%%%%%%%%%%%%%%%%%%%%%%%

\section{Experimental results}

%%%%%%%%%%%%%%%%%%%%%%%%%%%%%%%%%%%%%%%%%%

\subsection{AC Susceptibility in Zero Field}
\label{susceptibility}

\begin{figure}
	\includegraphics[width=1.0\linewidth]{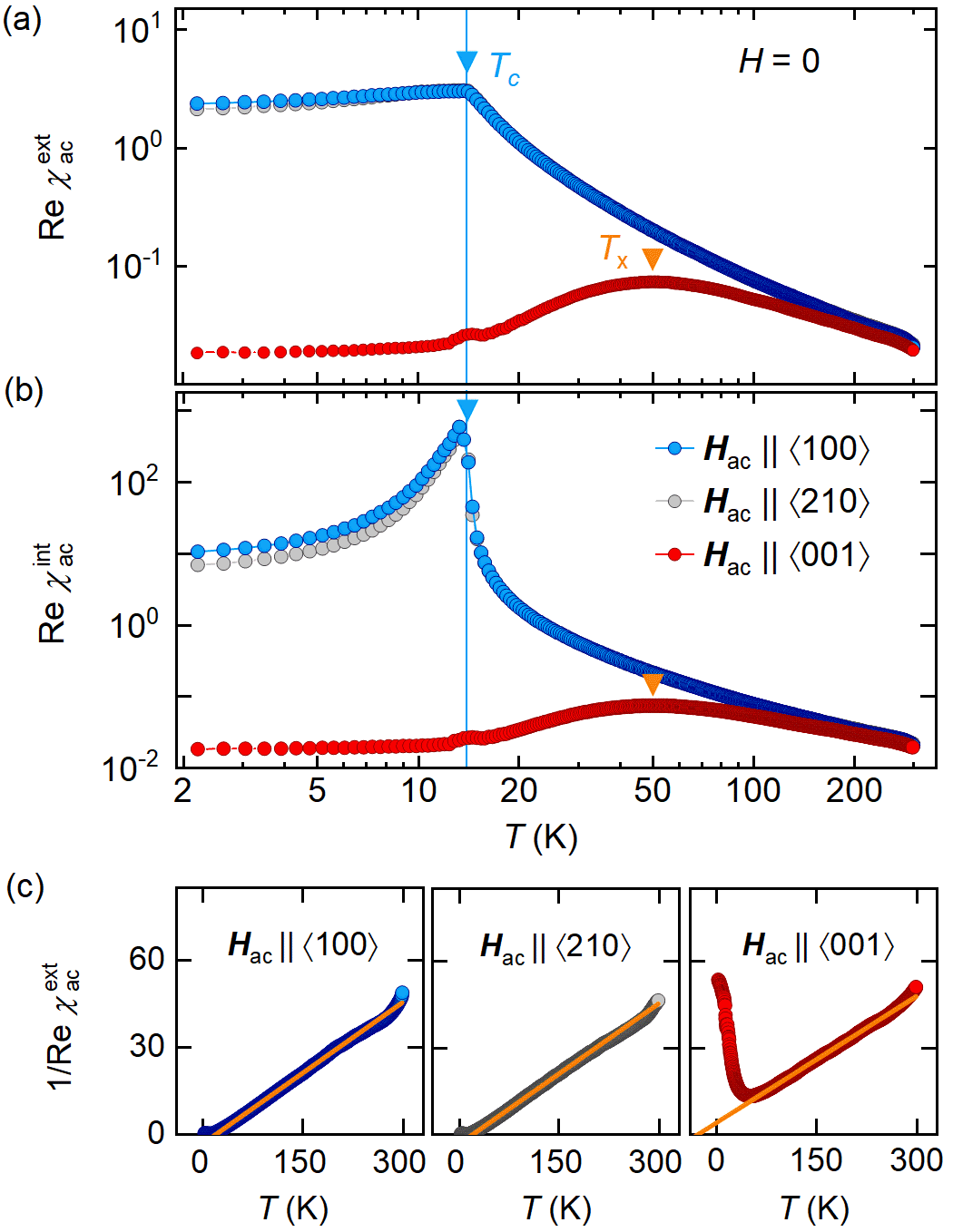}
	\caption{\label{figure1}Real part of the ac susceptibility as a function of temperature in zero magnetic field, $H = 0$. Data are shown for ac excitation fields applied along the major crystallographic axes $\langle100\rangle$ (blue symbols), $\langle210\rangle$ (gray symbols), and $\langle001\rangle$ (red symbols). (a)~Measured susceptibility on a double-logarithmic scale. (b)~Susceptibility on a double-logarithmic scale when correcting for the demagnetization effects in the ac excitation field. (c)~Inverse of the measured ac susceptibility. Orange lines are linear Curie--Weiss fits.}
\end{figure}

\replaced[id=CP]{Shown in Fig.~\ref{figure1}(a) is}{The presentation of our experimental results starts with} the temperature dependence of the ac susceptibility at zero field for excitation fields parallel to the major crystallographic axes\deleted[id=CP]{, shown in Fig.~\ref{figure1}(a)}. As a function of decreasing temperature the susceptibility exhibits a Curie--Weiss dependence. In the basal plane, this behavior continues until a kink at $T_{c} = 14$~K is followed by a decrease of the susceptibility, characteristic of the onset of long-range magnetic order. Data for excitations along $\langle100\rangle$ and $\langle210\rangle$ essentially track each other, characteristic of a vanishingly small in-plane anisotropy. 

For $\langle001\rangle$, the absolute value of the susceptibility is smaller, characteristic of a hard magnetic axis, and the temperature dependence deviates from the Curie--Weiss-like behavior observed at the highest temperatures studied. Instead, a shallow maximum may be observed at $T_{x} \approx 50$~K. \added[id=CP]{The presence of a broad maximum of the susceptibility in the paramagnetic state is ubiquitous in many materials and may be due to a range of different mechanisms. For instance, in metallic systems it may be related to a maximum in the density of states, while systems lacking inversion symmetry may feature the emergence of components of the spin fluctuations at finite wavevectors~\cite{Ishikawa_1985}.}

\replaced[id=CP]{In ErB$_2$ the}{Such a} maximum is characteristic of a crossover that may be attributed to the emergence of strong magnetic anisotropy at low temperatures \added[id=CP]{imposed by CEFs as discussed below}. Below $T_{x}$, the magnetic anisotropy prevails over thermal fluctuations, effectively reducing the dimensionality of the magnetic fluctuation spectrum by restricting fluctuations to the hexagonal basal plane. At $T_{c}$, a weak cusp is observed that may reflect a small misalignment of the ac excitation field with respect to the $\langle001\rangle$ axis of the order of $1^\circ$, resulting in a tiny in-plane component of the excitation field.

When calculating the ac susceptibility with respect to the internal magnetic field scale by taking into account the demagnetization effects, qualitatively similar behavior is observed, as shown in Fig.~\ref{figure1}(b). However, the absolute value of the susceptibility increases, in particular for temperatures around $T_{c}$ and below. High values of up to 600 are reached, reminiscent of soft ferromagnets.

For further quantitative analysis, the inverse of the measured ac susceptibility in the paramagnetic regime is fitted \replaced[id=CP]{with a}{linear} Curie--Weiss dependence, as shown in Fig.~\ref{figure1}(c). In the paramagnetic state, the susceptibility is comparably small and demagnetization effects are mostly negligible. For excitation fields within the basal plane, a linear regression between 50~K and 250~K yields fluctuating magnetic moments of $\mu_{\mathrm{eff}}^{\langle100\rangle} = 9.13~\mu_{\mathrm{B}}\,\mathrm{f.u.}^{-1}$ and $\mu_{\mathrm{eff}}^{\langle210\rangle} = 9.15~\mu_{\mathrm{B}}\,\mathrm{f.u.}^{-1}$ as well as Curie--Weiss temperatures $\mathit{\Theta}_{\mathrm{CW}}^{\langle100\rangle} = \mathit{\Theta}_{\mathrm{CW}}^{\langle210\rangle} = 19.0$~K. For excitation fields parallel to $\langle001\rangle$, a linear regression between 100~K and 250~K yields $\mu_{\mathrm{eff}}^{\langle001\rangle} = 9.64~\mu_{\mathrm{B}}\,\mathrm{f.u.}^{-1}$ and $\mathit{\Theta}_{\mathrm{CW}}^{\langle001\rangle} = -29.8$~K. The fluctuating moments are close to the free-ion value of Er$^{3+}$, $\mu_{\mathrm{eff}} = 9.58~\mu_{\mathrm{B}}\,\mathrm{f.u.}^{-1}$, consistent with magnetism that is carried by the 4f electrons on the rare-earth sites. \replaced[id=CR]{The positive $\mathit{\Theta}_{\mathrm{CW}}$ for fields within the basal plane suggests ferromagnetic interactions, whereas the negative $\mathit{\Theta}_{\mathrm{CW}}$ along $\langle001\rangle$ may reflect antiferromagnetic interactions or, more likely, crystal-field level repopulation~\cite{Xing_2011}.}{The positive and the negative sign of $\mathit{\Theta}_{\mathrm{CW}}$ in the basal plane and perpendicular to the basal plane, respectively, suggest that the interactions are ferromagnetic within the basal plane and antiferromagnetic perpendicular to it.} \added{Curie-Weiss parameters inferred from the dc susceptibility at $0.1$~T (not shown) are consistent with those obtained from the ac susceptibility .}

Antiferromagnetic interactions in a hexagonal environment as well as competing interactions in general may give rise to frustration effects. The ratio $f = -\mathit{\Theta}_{\mathrm{CW}}/T_{c}$ is commonly used to quantify the suppression of long-range magnetic order due to frustration, where values exceeding ten are considered a hallmark of strongly frustrated systems~\cite{1994_Ramirez_AnnuRevMaterSci}. In ErB$_{2}$, a value of $f_{\langle001\rangle} = 2.1$ is observed for the c-axis, characteristic of a system in which very weak geometrical frustration may be present.

%%%%%%%%%%%%%%%%%%%%%%%%%%%%%%%%%%%%%%%%%%

\subsection{Specific Heat}
\label{specificheat}

\begin{figure}
	\includegraphics[width=1.0\linewidth]{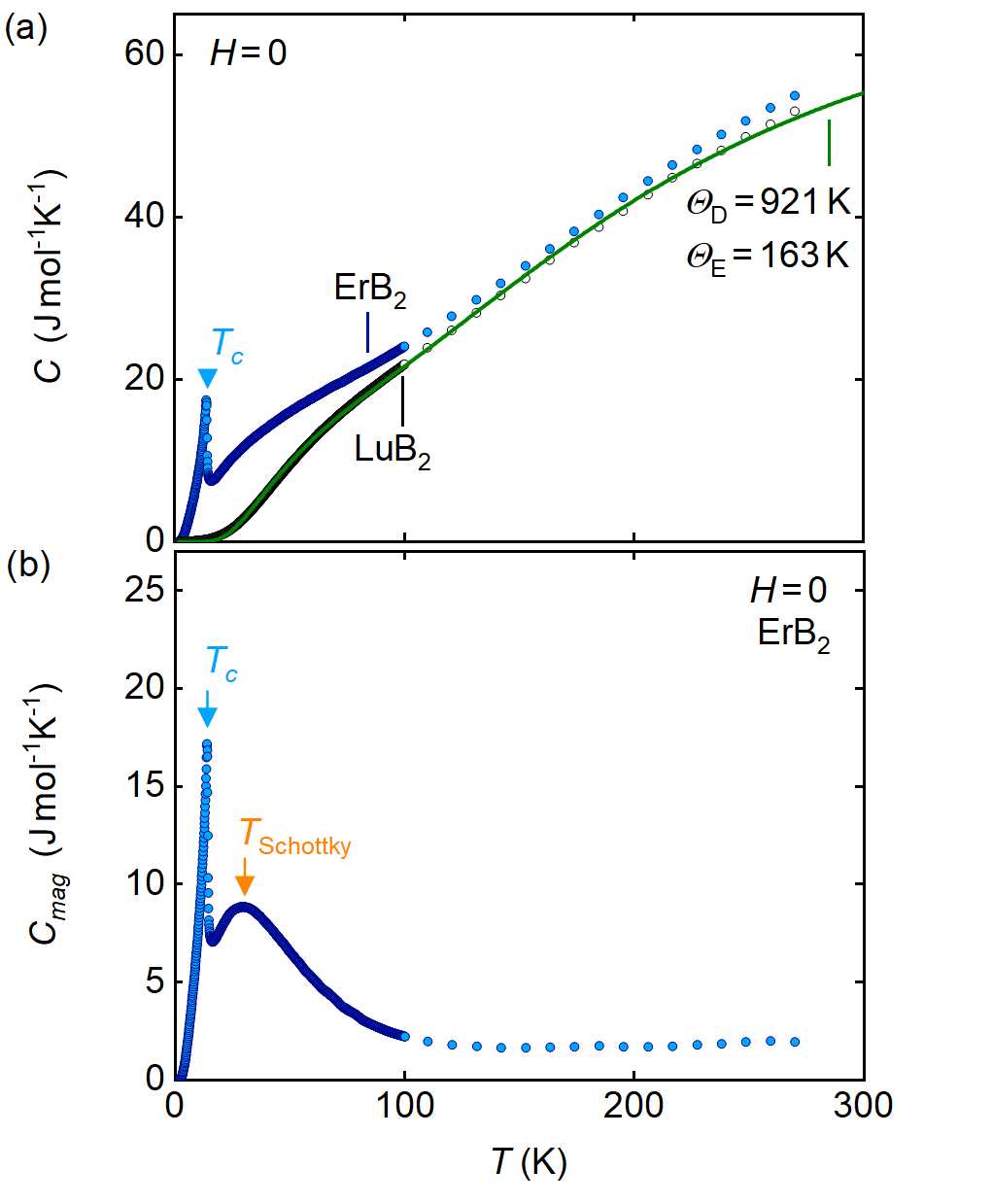}
	\caption{\label{figure2}\added{Specific heat of ErB$_{2}$ as a function of temperature in zero magnetic field.} (a) Specific heat of ErB$_2$ shown alongside data of nonmagnetic LuB$_{2}$ (open symbols) and a \replaced{combined Debye-Einstein model with Debye and Einstein temperatures $\mathit{\Theta}_{\mathrm{D}} = 921$~K and $\mathit{\Theta}_{\mathrm{E}} = 163$~K (solid green line)}{Debye model with a Debye temperature $\mathit{\Theta}_{\mathrm{D}} = 735$~K} for comparison. (b) Magnetic contribution to the specific heat\replaced{}{ divided by temperature} after subtracting the specific heat of LuB$_{2}$ as nonmagnetic contribution. \replaced{An orange arrow marks the Schottky anomaly.}{Inset: Enlarged view at low temperatures. (c)~Magnetic contribution to the entropy, approaching $4\ln2$ at high temperatures. Inset: Enlarged view at low temperatures.}}
\end{figure}

\begin{figure}
	\includegraphics[width=1.0\linewidth]{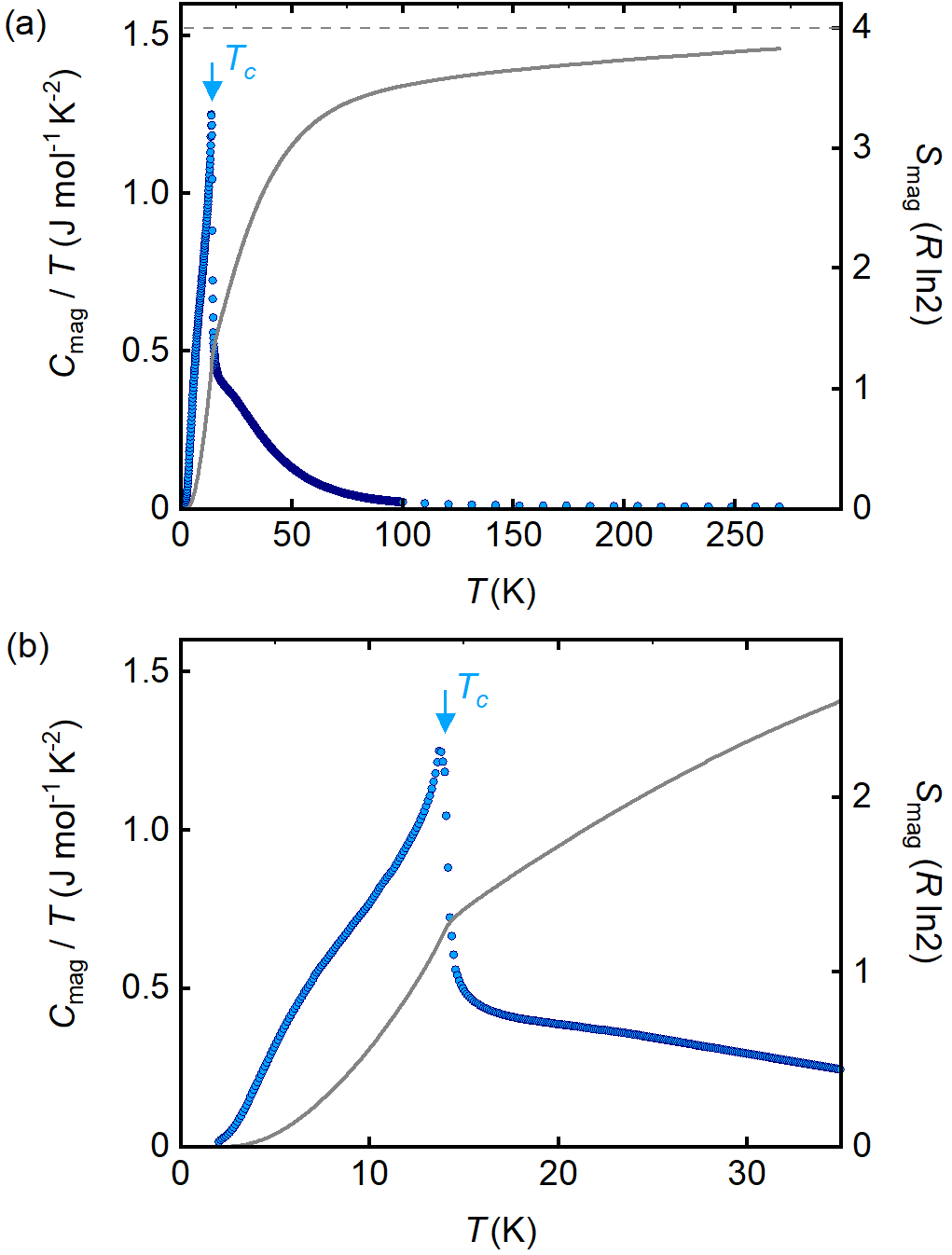}
	\caption{\label{figure3}\added{Magnetic entropy of ErB$_2$. (a)~Magnetic entropy  shown together with the magnetic contribution to the specific heat divided by temperature. The former approaching $4R$ln2 (dashed line) at high temperatures. (b) Enlarged view at low temperatures. The magnetic entropy released at $T_c$ is $1.3R$ln2.}}
\end{figure}

The specific heat as a function of temperature is shown in Fig.~\ref{figure2}(a). At low temperatures, a  lambda anomaly is observed at $T_{c}$, characteristic of a second-order phase transition consistent with the susceptibility. At high temperatures, the specific heat slowly approaches the Dulong--Petit value of $9R = 74.8~\mathrm{J\,mol}^{-1}\mathrm{K}^{-1}$. To account for nonmagnetic contributions, the specific heat of the isostructural C32 compound  LuB$_{2}$ was measured on a polycrystal prepared by means of arc melting~\cite{2016_Bauer_RevSciInstrum}. At high temperatures, the specific heat of nonmagnetic LuB$_{2}$ essentially tracks that of ErB$_{2}$ at slightly smaller absolute values. Below about 100~K, ErB$_{2}$ exhibits additional contributions that are absent in its nonmagnetic counterpart. \replaced{}{These contributions may be attributed to magnetic fluctuations that emerge well above the onset of long-range magnetic order at $T_{c}$. They relate to the crossover at $T_{x}$ observed in the ac susceptibility.}

For comparison, a phonon contribution is shown as calculated by means of \replaced{a combined Debye-Einstein model}{the Debye model} (solid line). \replaced{Debye and Einstein temperatures $\mathit{\Theta}_{\mathrm{D}} = 921$~K and $\mathit{\Theta}_{\mathrm{E}} = 163$~K and mode fractions $\mathit{n}_{\mathrm{D}} = 0.728$ and $\mathit{n}_{\mathrm{E}} = 0.272$ are inferred from a fit to the specific heat of LuB$_2$. The data are in excellent agreement with the model over the whole temperature range studied, indicating that both acoustic and optical phonon modes contribute significantly to the specific heat of LuB$_2$ and ErB$_2$.}{The Debye temperature $\mathit{\Theta}_{\mathrm{D}} = 735$~K was chosen such that the calculated phonon contribution approaches the data at high temperatures while staying at smaller absolute values. In particular at intermediate temperatures, namely between 30~K and 180~K, a large discrepancy is observed between the Debye model and the measured specific heats of both LuB$_{2}$ and ErB$_{2}$. Akin to other diborides~\cite{2009_Mori_PhysRevB, 2008_Novikov_InorgMater, 2014_Bauer_PhysRevB}, discrepancies with nonmagnetic LuB$_{2}$ may be attributed to the reduced effective dimensionality associated with the layered crystal structure, beyond the assumptions of the Debye model~\footnote{At low temperatures, typically $T < \mathit{\Theta}_{\mathrm{D}}/10$, the Debye model yields a phonon contribution of the specific heat that scales with temperature to the power of three. At high temperatures, i.e.,  $T > \mathit{\Theta}_{\mathrm{D}}$, the Debye model approaches the Dulong--Petit value. At intermediate temperatures, the Debye model provides a rough approximation of the phonon contribution to the specific heat that describes the situation in many materials surprisingly well. The deviations observed in the diborides do not permit to infer a Debye temperature from the low-temperature part of the Debye model using a linear fit to $C/T$ as a function of $T^{2}$ for temperatures $T_{c} < T < \mathit{\Theta}_{\mathrm{D}}/10$.}} 

Using the specific heat of LuB$_{2}$ to subtract nonmagnetic signal contributions, the magnetic part of the specific heat \replaced{of ErB$_2$}{divided by temperature} is shown in Fig.~\ref{figure2}(b). This contribution is dominated by the lambda anomaly at $T_{c}$ \replaced{and a Schottky anomaly centered at $T_\mathrm{Schottky} = 30$~K, corresponding to a CEF splitting of the ground-state multiplet into Kramers doublets. In a two-level Schottky model involving two doublets, the position of the Schottky maximum corresponds to an energy splitting of approximately 70~K (6~meV) between the ground state and the first excited doublet. In light of this CEF splitting, the temperature-dependent magnetic anisotropy observed in ac susceptibility measurements can be attributed to the thermal population of excited CEF levels.}{Below $T_{c}$, an additional hump may be observed around 8~K, reminiscent to the behavior observed in the itinerant antiferromagnet CrB$_{2}$~\cite{2014_Bauer_PhysRevB}.}

\added{Figure~\ref{figure3}(a) shows the magnetic contribution to the specific heat  of ErB$_2$ divided by temperature. Numerically integrating $C_\mathrm{mag}/T$ yields the magnetic entropy, also shown in Fig.~\ref{figure3}(a). Above $T_{c}$, sizable magnetic contributions to both the specific heat and the entropy are observed. At high temperatures the entropy gradually approaches the value of $4R$ln2, in excellent agreement with magnetism arising due to Er$^{3+}$ moments with $J=15/2$.}

\added{Figure~\ref{figure3}(b) shows an enlarged view of $C_\mathrm{mag}/T$ and $S_\mathrm{mag}$ at low temperatures. Below the lambda anomaly at $T_{c}$, an additional hump may be observed around 8~K, reminiscent of the behavior observed in the itinerant antiferromagnet CrB$_{2}$~\cite{2014_Bauer_PhysRevB}. }
A reliable extrapolation of the specific heat to zero temperature is not possible due to the large slope of $C/T$ at the lowest temperatures studied. However, the value of $20~\mathrm{mJ\,mol}^{-1}\mathrm{K}^{-2}$ at 2~K implies a Sommerfeld coefficient of the order of a few $\mathrm{mJ\,mol}^{-1}\mathrm{K}^{-2}$ in ErB$_{2}$, characteristic of a material with comparatively weak electronic correlations. \replaced{}{At temperatures above $T_{c}$, sizeable magnetic contributions are observed as associated with the crossover phenomenon at $T_{x}$, consistent with the ac susceptibility. Numerically integrating the specific heat divided by temperatures yields the magnetic contribution to the entropy shown in Figs.~\ref{figure2}(c). For Er$^{3+}$, Hund's rules imply $J = 15/2$ and a magnetic contribution to the entropy $S_\mathrm{mag}$ = $4R\ln2$ at high temperatures.}  The magnetic phase transition at $T_{c}$ is associated with a clear kink \added{in the magnetic entropy}, corresponding to the release of approximately $1.3R\ln 2$ at $T_{c}$.\replaced[id=CR]{ This value slightly exceeds the $R\ln2$ expected for a single Kramers doublet, suggesting a partial thermal population of the first excited CEF level.}{This indicates that the anisotropic magnetic fluctuations in the paramagnetic regime above $T_{c}$ carry large parts of the magnetic entropy.Up to room temperature, the magnetic contribution to the entropy approaches $4R\ln2$, in excellent agreement with magnetism arising due to Er$^{3+}$ moments.} 

%%%%%%%%%%%%%%%%%%%%%%%%%%%%%%%%%%%%%%%%%%

\subsection{Magnetization and AC Susceptibility}
\label{magnetization}

\begin{figure}
	\includegraphics[width=1.0\linewidth]{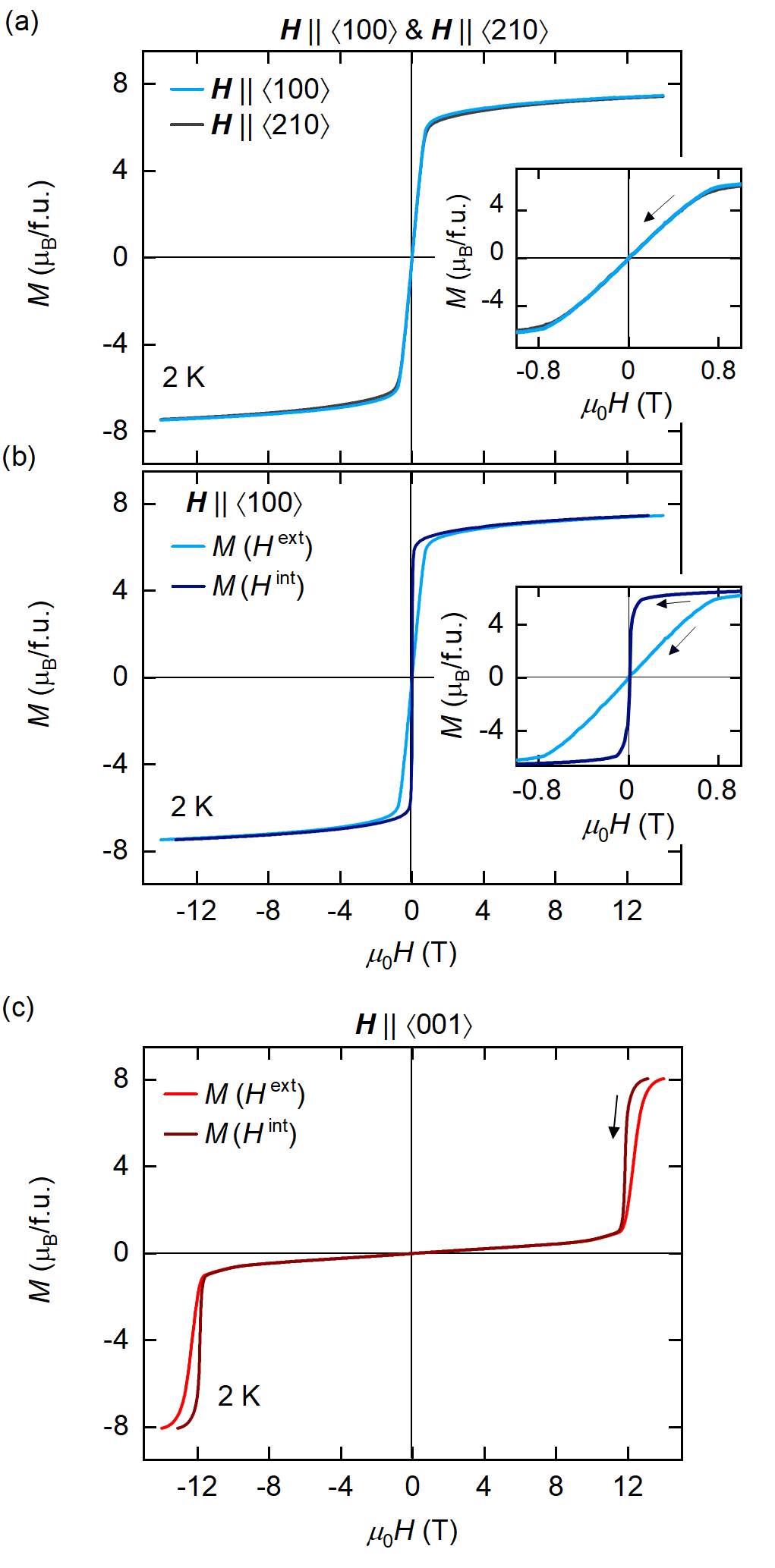}
	\caption{\label{figure4}Magnetization as a function of field at low temperature. (a)~Magnetization for field along $\langle100\rangle$ and $\langle210\rangle$. Data are shown for decreasing field (black arrow). (b)~Magnetization for field along $\langle100\rangle$ as a function of applied and internal field. Insets: Enlarged view of the low-field behavior. (c)~Magnetization for field along $\langle001\rangle$ as a function of applied and internal field.}
\end{figure}

The evolution of the magnetic properties as a function of field is illustrated by means of the field dependence of the magnetization. For magnetic field applied within the basal plane, shown in Fig.~\ref{figure4}(a), the magnetization exhibits an S-shape with pronounced changes of slope around $\pm0.8$~T characteristic of the onset of the field-polarized state at high fields. Up to 14~T\deleted[id=CP]{, the highest field studied,} the slope remains finite with the magnetization reaching $7.5~\mu_{\mathrm{B}}/\mathrm{f.u.}$. This value is smaller than the free-ion value of $9.0~\mu_{\mathrm{B}}\,\mathrm{f.u.}^{-1}$ of Er$^{3+}$. \replaced{}{In combination with the lack of saturation, such discrepancy typically is attributed to itinerant magnetism, suggesting that the 4f electrons of Er$^{3+}$ in fact may be hybridized with d electrons at the Fermi level.} 

Data for field parallel to $\langle100\rangle$ and $\langle210\rangle$ essentially track each other, characteristic of vanishing in-plane anisotropy. No hysteresis was observed at the level of the resolution of our study. 

When taking into account demagnetization effects, shown in Fig.~\ref{figure4}(b), the slope around zero field becomes very steep and the onset of the field-polarized regime shifts to much smaller field values. This behavior is strongly reminiscent of a soft ferromagnetic material in which magnetic domains reorient or change population under small magnetic fields.

For magnetic field parallel to $\langle001\rangle$, shown in Fig.~\ref{figure4}(c), the magnetization at small fields increases only weakly, staying below $1~\mu_{\mathrm{B}}\,\mathrm{f.u.}^{-1}$ up to 11~T. Around 12~T a pronounced increase is observed, characteristic of a spin-flip transition into the field-polarized state. The slope at high fields decreases, but remains finite with the magnetization reaching $8.1~\mu_{\mathrm{B}}\,\mathrm{f.u.}^{-1}$ at 14~T below the free-ion value of Er$^{3+}$. Taking into account demagnetization effects, the spin-flip transition is shifted to slightly smaller internal field values and the increase of the magnetization associated with the transition becomes very steep. 

\begin{figure}
	\includegraphics[width=1.0\linewidth]{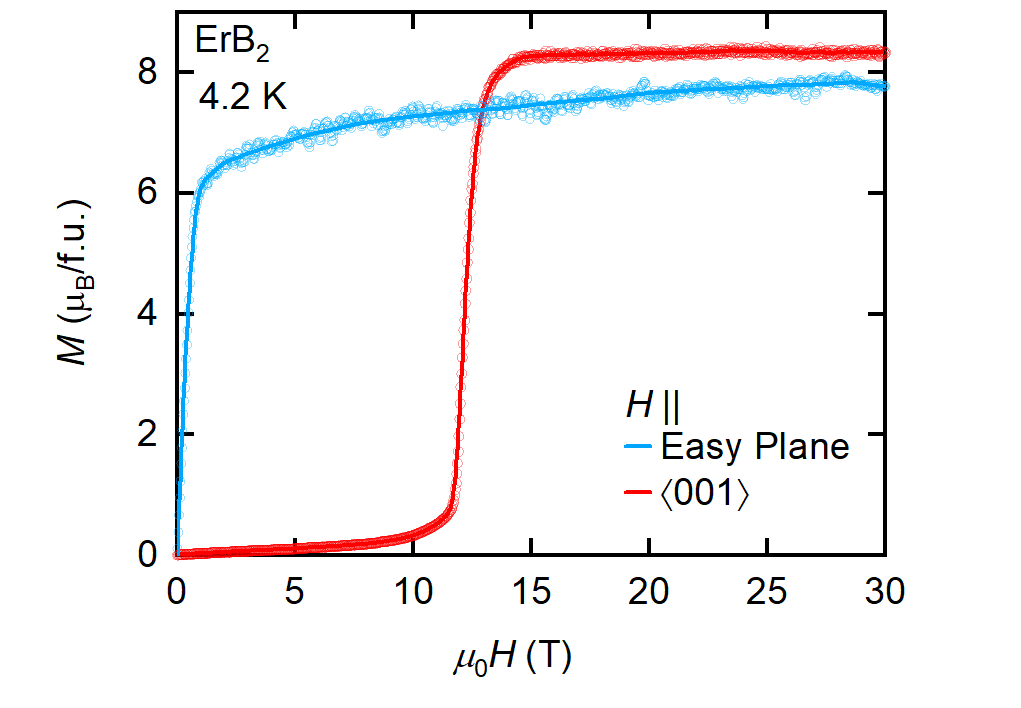}
	\caption{\label{figure5} \added{Magnetization at 4.2~K as a function of field applied within the basal plane and along the $\langle001\rangle$ direction as measured at HFML-FELIX. Data (open circles) were scaled to match the absolute values of the PPMS data and smoothed (solid line).}}
\end{figure}

\added{Data recorded at 4.2\,K for fields up to 30\,T applied parallel to $\langle001\rangle$ and within the basal plane are shown in Fig.~\ref{figure5}. For magnetic fields above $\approx$ 15~T parallel to $\langle001\rangle$ the magnetization flattens and reaches a value of $8.3~\mu_{\mathrm{B}}\,\mathrm{f.u.}^{-1}$ at the highest fields studied, confirming the spin-flip transition into the field-polarized state. The magnetization for field applied in plane retains a finite slope up to the highest fields studied, reaching a maximum value of $7.8~\mu_{\mathrm{B}}\,\mathrm{f.u.}^{-1}$, below the free ion value and the value for field applied along $\langle001\rangle$.} 

\deleted[id=CP]{Taken together, the magnetization for fields applied along the major crystallographic axes establish ErB$_{2}$ as a ferromagnetic material that is dominated by strong magnetic anisotropies, where $\langle001\rangle$ is the hard magnetic axis and the basal plane is magnetically easy with vanishingly small in-plane anisotropy.}

\deleted[id=CP]{The discrepancy between the observed saturation magnetization and the free-ion value of Er$^{3+}$, along with the lack of in-plane saturation, can in principle be attributed to several mechanisms, including mixed valency of the erbium ions, itinerant magnetism arising from 4$f$--5$d$ hybridization, or CEF effects. While mixed valency is well documented in various rare-earth-based compounds, such as the cerium-based Ce(Pd$_{1-x}$Rh$_x$)$_3$ system \cite{1981_Mihalisin_PhysRevLet} and the europium-based EuCu$_2$Si$_2$ \cite{1980_Hatwar_SolidStateCommunicationsa}, reports of mixed valency in erbium-based compounds appear less prevalent in the literature. In contrast, pronounced CEF effects are commonly observed in erbium-based 4$f$ systems \cite{2004_Watanuki_PhysRevBb, _Soderholm_b}. Considering the pronounced magnetic anisotropy and the Schottky-type anomaly in the specific heat (see Section~\ref{specificheat}), a CEF-induced splitting of the Hund’s rule ground state emerges as the most likely explanation. In this scenario, the lowest CEF level dominates the magnetic response at low temperatures. The reduction in the observed magnetic moment and its anisotropic nature can be attributed to the reduced moment and the anisotropic $g$-tensor associated with this ground state. Furthermore, the absence of saturation at high in-plane fields may be explained by a Van Vleck contribution, resulting from a mixing of excited states into the ground state.}

\begin{figure}
	\includegraphics[width=1.0\linewidth]{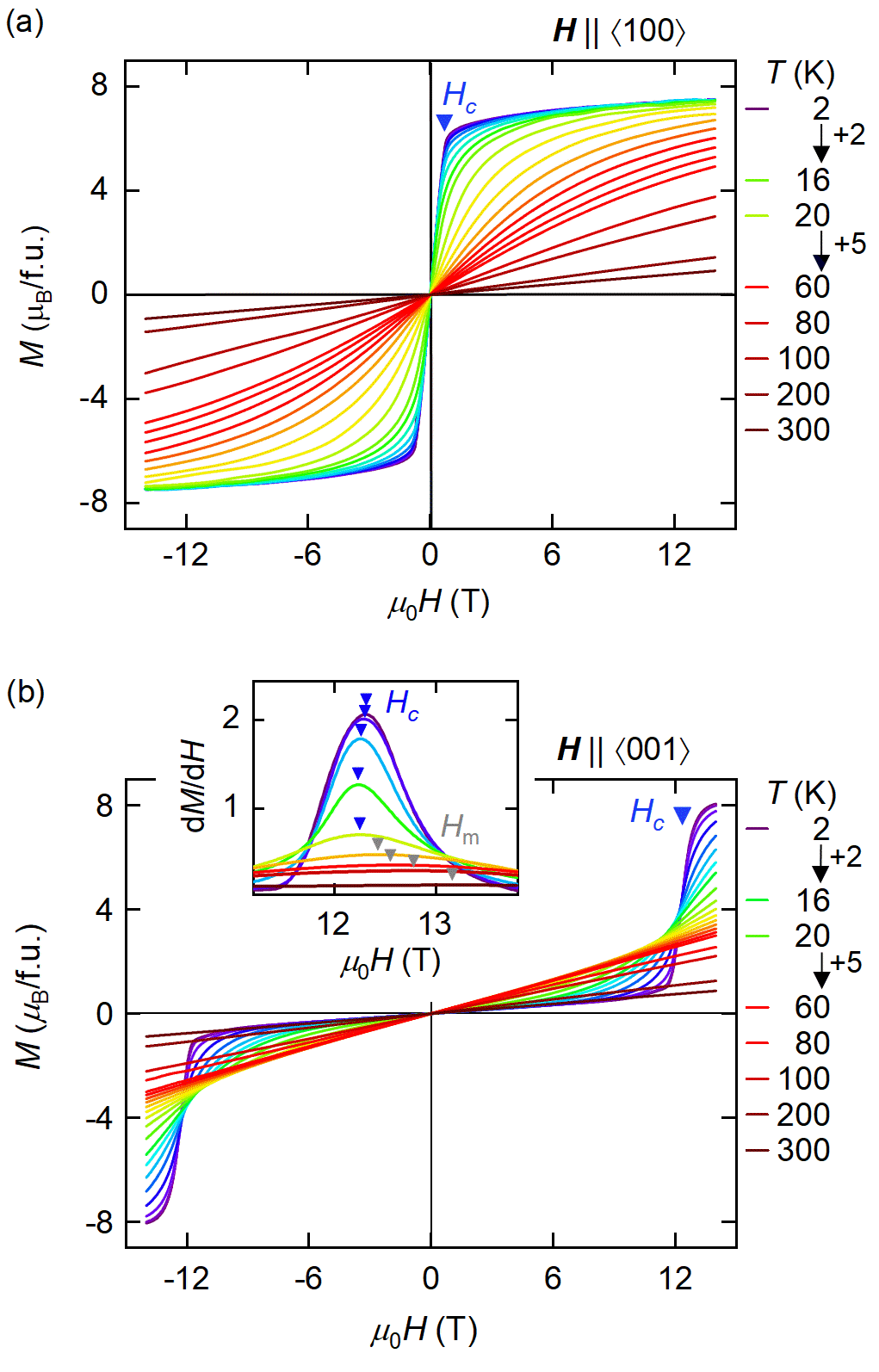}
	\caption{\label{figure6}Magnetization as a function of field for different temperatures. (a)~Magnetization for field along the magnetic easy $\langle100\rangle$ axis. (b)~Magnetization for field along the magnetic hard $\langle001\rangle$ axis. Inset: Differential susceptibility, $\mathrm{d}M/\mathrm{d}H$, around the spin-flip transition.}
\end{figure}

Under increasing temperature, the magnetization for field applied along the magnetic easy $\langle100\rangle$ axis remains qualitatively unchanged, as shown in Fig.~\ref{figure6}(a). The absolute value decreases monotonically as the change of slope associated with the onset of the field-polarized state at high fields becomes increasingly rounded, in particular for temperatures exceeding $T_{c}$. The characteristic field $H_{c}^{\langle100\rangle}$ may be tracked in terms of a point of inflection in the differential susceptibility (not shown). At temperatures well above $T_{c}$, the magnetization curve exhibits the shape of the Brillouin function, characteristic of paramagnetic behavior.

For field along the $\langle001\rangle$ hard magnetic axis shown in Fig.~\ref{figure6}(b), the jump-like increase of the magnetization around 12~T represents the most prominent feature, characteristic of a spin-flip into the field-polarized state. Under increasing temperature, this jump broadens and decreases. The transition field of the spin-flip, $H_{c}^{\langle001\rangle}$, may be inferred from the point of inflection in the magnetization, which may be tracked conveniently in the differential susceptibility, $\mathrm{d}M/\mathrm{d}H$ shown in the inset of Fig.~\ref{figure6}(b). Below $T_{c}$, a maximum is observed in $\mathrm{d}M/\mathrm{d}H$ that becomes less pronounced, \replaced[id=CR]{while}{where} the field value does not change for increasing temperature. Above $T_{c}$, a shallow maximum at $H_{m}$ may be resolved at least up to 30~K, which slowly shifts to larger field values. This maximum is associated with the crossover into the field-polarized state at high fields.

\begin{figure}
	\includegraphics[width=1.0\linewidth]{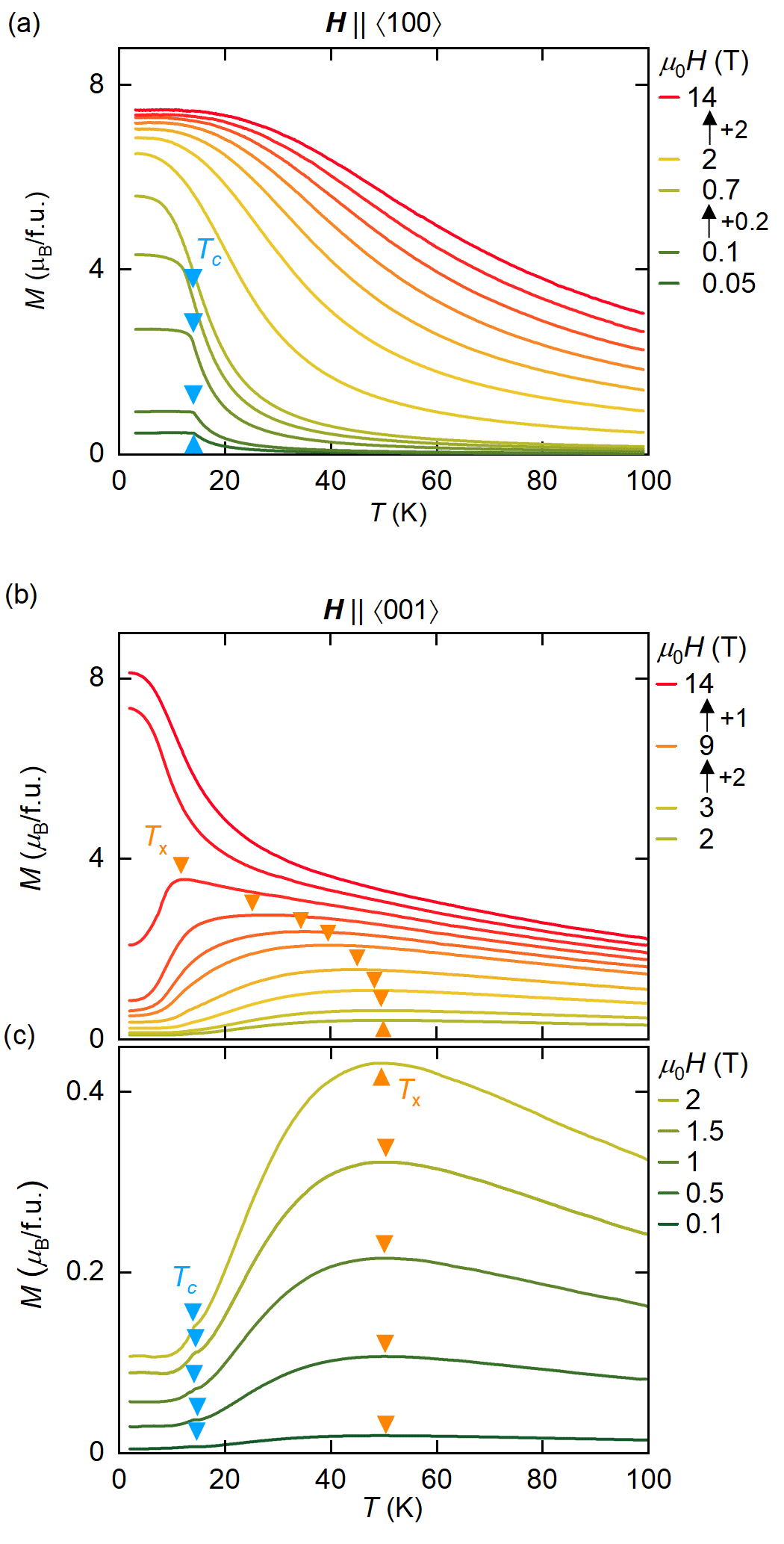}
	\caption{\label{figure7}Magnetization as a function of temperature for different applied magnetic fields. (a)~Magnetization for field along the magnetic easy $\langle100\rangle$ axis. \mbox{(b),(c)}~Magnetization for field along the magnetic hard $\langle001\rangle$ axis.}
\end{figure}

The magnetization as a function of the temperature provides complementary information to the magnetic field dependence. For field along the magnetic easy $\langle100\rangle$ axis shown in Fig.~\ref{figure7}(a), the magnetization at the lowest fields essentially corresponds to the ac susceptibility. As a function of decreasing temperature, a Curie--Weiss-dependence of the magnetization is followed by a kink at $T_{c} = 14$~K and becomes essentially constant below $T_{c}$. Under increasing field, the absolute value of the magnetization increases, the kink at $T_{c}$ gets rounded and shifts to smaller temperatures. At high fields, the magnetization increases smoothly and levels off at values of up to $7.5~\mu_{\mathrm{B}}/\mathrm{f.u.}$ at low temperatures. As emphasized above, internal field values at small fields are significantly smaller than the applied field values and, perhaps more importantly, changes as a function of temperature as combined with the strength of the demagnetization effects. Therefore, the kink at $T_{c}$ under finite magnetic field may be attributed to domain reorientation or repopulation of multi-domain ferromagnetic order.

For field along the $\langle001\rangle$ hard magnetic axis, the temperature dependence of the magnetization is shown in Fig.~\ref{figure7}(b) and Fig.~\ref{figure7}(c) for high and low fields, respectively. At the lowest fields, the magnetization resembles the ac susceptibility, namely a shallow maximum is observed around $T_{x} = 50$~K, followed by a weak cusp at $T_{c}$ that may be associated with a fortuitous small in-plane component of the applied magnetic field. Under increasing field, the absolute value of the magnetization increases. The temperature of the cusp remains essentially constant and vanishes for fields exceeding ${\sim}2$~T. The maximum at $T_{x}$ becomes more pronounced and shifts to lower temperatures, vanishing above 12~T. At \added[id=CR]{the} highest fields, the magnetization exhibits high values at low temperatures, reaching more than $8~\mu_{\mathrm{B}}/\mathrm{f.u.}$, characteristic of the field-polarized regime. These values are in excellent agreement with the data as a function of field.

\begin{figure}
	\includegraphics[width=1.0\linewidth]{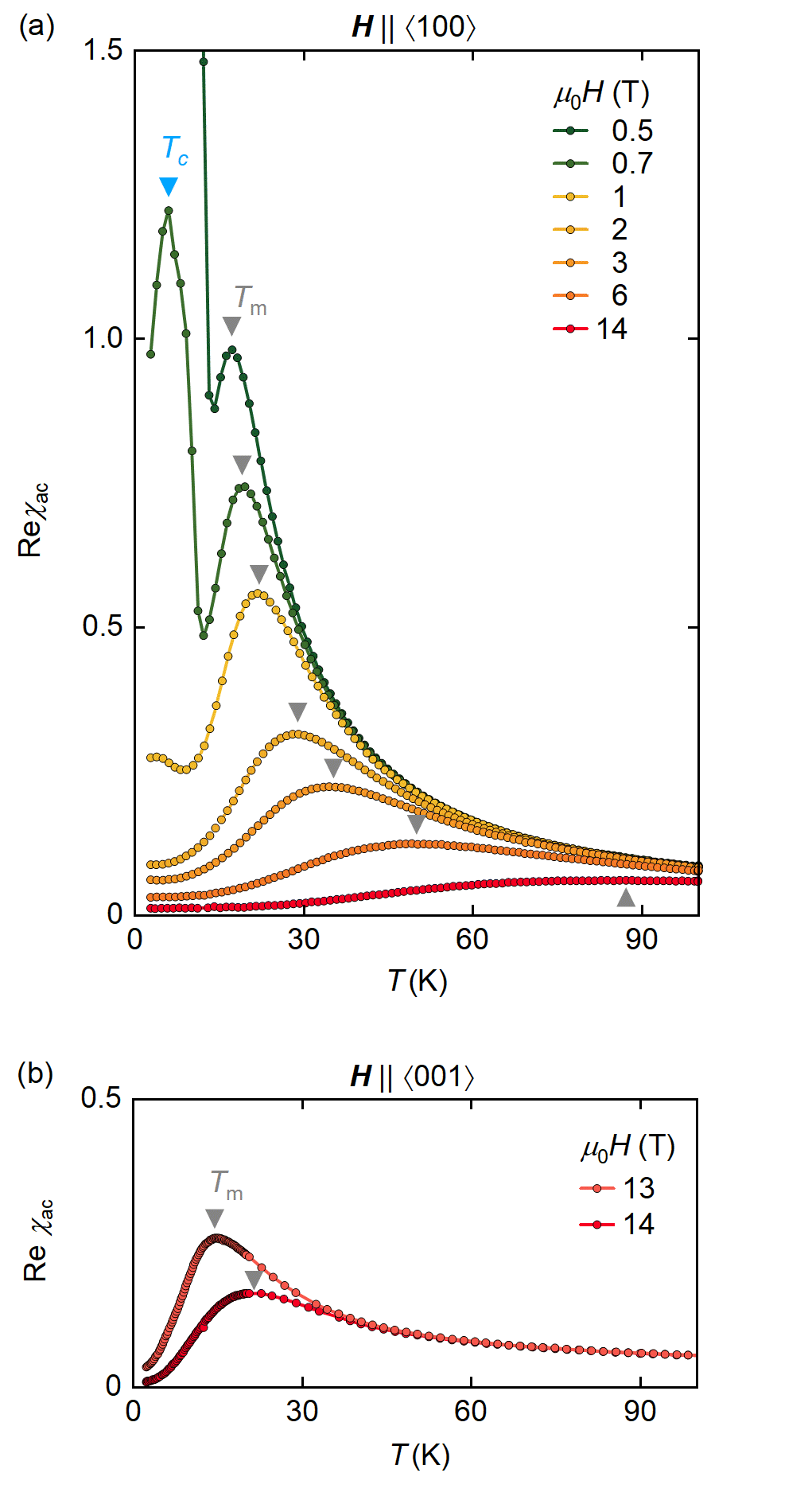}
	\caption{\label{figure8}AC susceptibility as a function of temperature for different applied magnetic fields. (a)~Susceptibility for field along the magnetic easy $\langle100\rangle$ axis. \mbox{(b),(c)}~Susceptibility for field along the magnetic hard $\langle001\rangle$ axis.}
\end{figure}

While the ac susceptibility at the lowest fields (not shown) mirrors the evolution of the magnetization under field, it exhibits an additional crossover at high fields. As shown in Fig.~\ref{figure8}(a) for field along the $\langle100\rangle$ easy magnetic axis, the temperature dependence of the ac susceptibility exhibits a broad maximum, denoted $T_{m}$, above the onset of ferromagnetic order at $T_{c}$. Under increasing field, the absolute value of the susceptibility decreases and the maximum shifts to higher temperatures. For field along the magnetic hard $\langle001\rangle$ axis, shown in Fig.~\ref{figure8}(b), a similar maximum is observed for fields above the spin-flip transition at 12~T. These maxima mark the crossover between the field-polarized regime at low temperatures and the paramagnetic regime at high temperatures, as reported in a wide range of magnetic materials~\cite{1997_Thessieu_JPhysCondensMatter, 2010_Bauer_PhysRevB, 2022_Bauer_PhysRevMater}.

\subsection{Electrical Resistivity}
\label{resistivity}

\begin{figure}
	\includegraphics[width=1.0\linewidth]{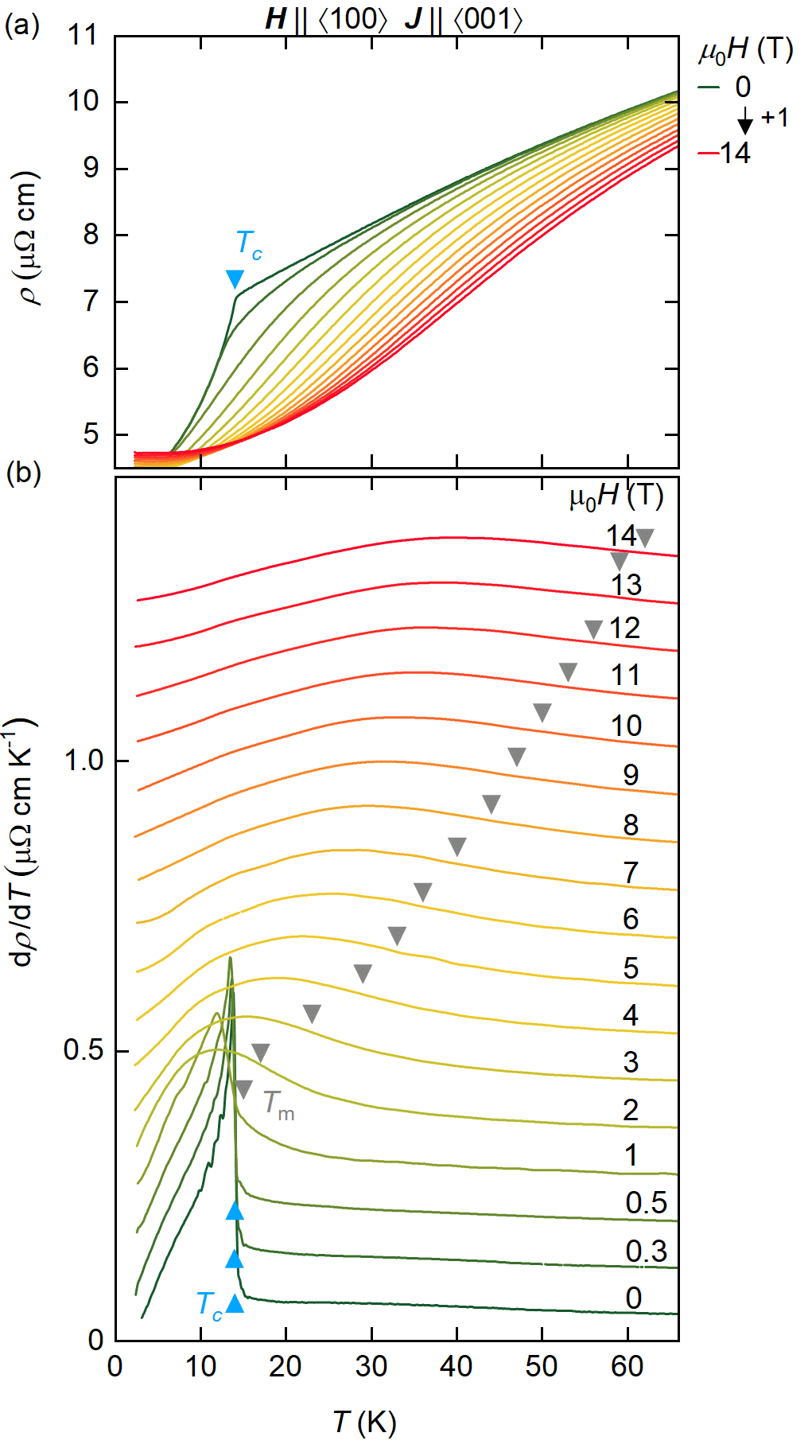}
	\caption{\label{figure9}Electrical resistivity as a function of temperature for magnetic field along $\langle100\rangle$ and current along $\langle001\rangle$. (a)~Resistivity for different magnetic fields. (b)~Derivative of the resistivity with respect to temperature for different magnetic fields.}
\end{figure}

\begin{figure}
	\includegraphics[width=1.0\linewidth]{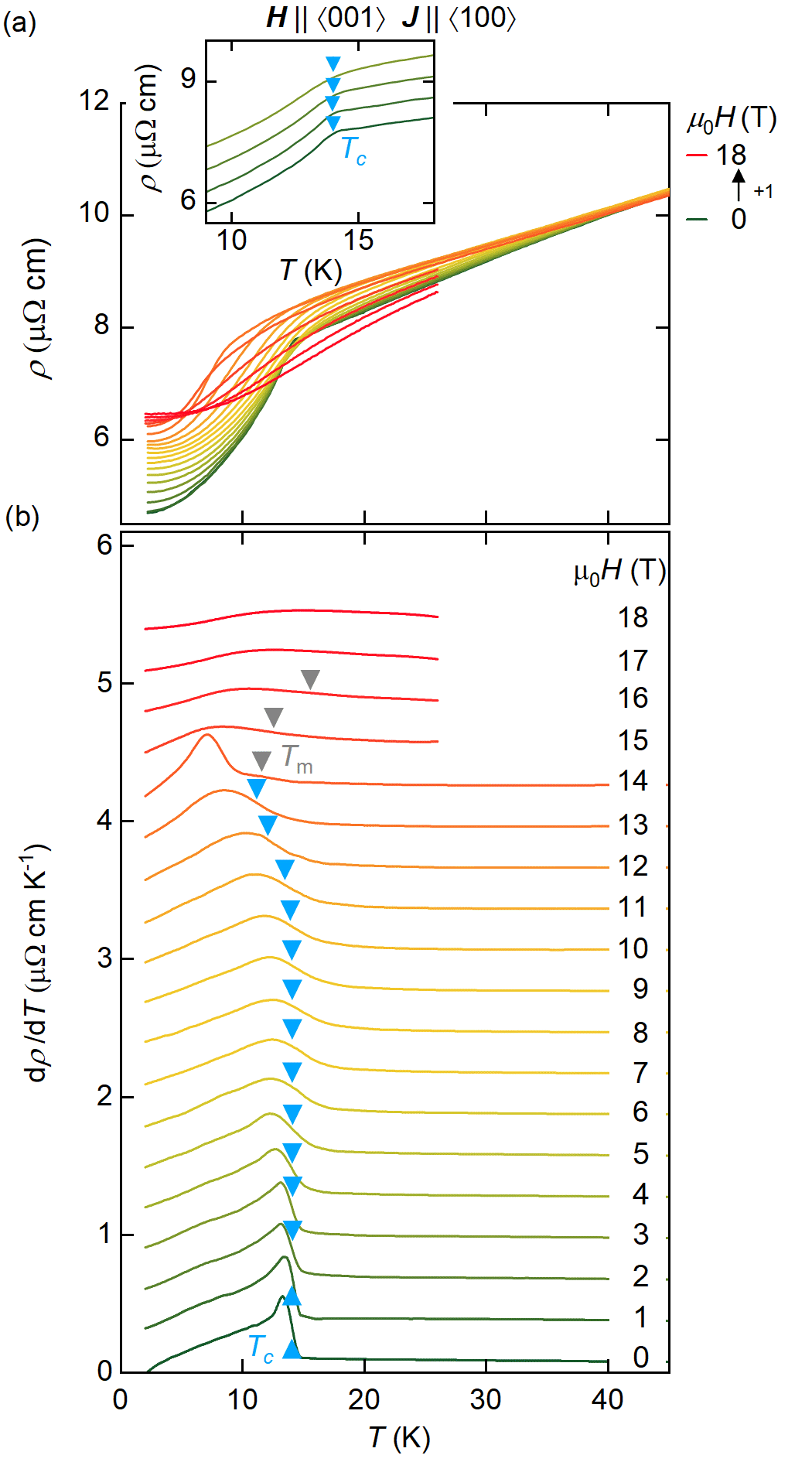}
	\caption{\label{figure10}Electrical resistivity as a function of temperature for magnetic field along $\langle001\rangle$ and current along $\langle100\rangle$. (a)~Resistivity for different magnetic fields. Inset: Enlarged view of the behavior at small fields around $T_{c}$. (b)~Derivative of the resistivity with respect to temperature for different magnetic fields.}
\end{figure}

The electrical resistivity was measured for two configurations: (i) magnetic field along an easy axis and current along the hard axis, shown in Fig.~\ref{figure9}, and, (ii), magnetic field along the hard axis and current along an easy axis, shown in Fig.~\ref{figure10}. As shown in Fig.~\ref{figure9}(a), for field along $\langle100\rangle$ and current along $\langle001\rangle$ the resistivity decreases with decreasing temperature characteristic of a good \deleted{a}metal. A kink at $T_{c}$ is followed by a pronounced decrease of the resistivity, characteristic of the onset of ferromagnetic order and a freezing out of a scattering mechanism, notably spin \replaced[id=CR]{fluctuations}{fluctations}. The temperature dependence of the resistivity below $T_{c}$ varies as $T^{2}$. Given the strongly \replaced{anisotropic}{anisotropy} ferromagnetic state it is not possible to discriminate electron-magnon scattering from electron-electron scattering expected in a Fermi liquid. Under small magnetic fields, the resistivity \added[id=CR]{curves} at low temperatures track each other up to $T_{c}$, where the kink is rounded and shifted to lower temperatures. When further increasing the applied magnetic field, the resistivity varies characteristic of a conventional magnetoresistance.

The evolution of the characteristic features under field associated with features in the magnetization, susceptibility and specific heat, may be tracked quite conveniently in the derivative of the resistivity with respect to temperature, $\mathrm{d}\rho_{xx}/\mathrm{d}T$, shown in Fig.~\ref{figure9}(b). At zero field, $\mathrm{d}\rho_{xx}/\mathrm{d}T$ is strongly reminiscent of the electronic contribution to the specific heat divided by temperature reported above. This establishes that both quantities are dominated by the same electronic density of states at the Fermi level. 

A point of inflection corresponds to the magnetic transition at $T_{c}$. Under increasing field, the asymmetric peak at $T_{c}$ decreases and broadens, shifting to lower temperatures. Above 0.5~T, the pronounced peak develops into a broad maximum that shifts to higher temperatures. The point of inflection at the high-temperature side, denoted $T_{m}$, coincides with the crossover between the field-polarized regime at low temperatures and the paramagnetic regime at high temperatures. In view of the discussion of the magnetization data in Figs.~\ref{figure4}(b) and \ref{figure7}(a), the kink at $T_{c}$ in finite applied field may be attributed to domain reorientation processes of multi-domain ferromagnetic order.

As shown in Fig.~\ref{figure10}(a), the resistivity for current along $\langle100\rangle$ in zero field resembles the resistivity for current along $\langle001\rangle$, including the absolute value, the pronounced kink at $T_{c}$, and the $T^{2}$ temperature dependence below $T_{c}$. For applied  magnetic field along $\langle001\rangle$, the kink at $T_{c}$ becomes rounded while staying at unchanged temperature. In general, positive magnetoresistance is observed up to 13~T when the magnetoresistance becomes negative for all but the lowest temperatures.

The derivative of the resistivity as a function of temperature, shown in Fig.~\ref{figure10}(b), allows to track the evolution of the onset of ferromagnetic order under field more conveniently. At zero field, the data resemble the behavior for current along $\langle001\rangle$ and in turn the electronic contribution to the specific heat divided by temperature. Under field, the maximum attributed to the onset of magnetic order becomes smaller, and the temperature of the point of inflection on the high-temperature side, marking $T_{c}$, remains unchanged up to 9~T. For larger fields, $T_{c}$ shifts to smaller temperatures before vanishing above 14~T. When taking into account demagnetizing fields, the maximum observed at 14~T may be attributed to a crossover emerging from the spin-flip transition. Above 14~T, a shallow maximum is observed that shifts to higher temperatures with increasing field. Similar to the properties for field along $\langle100\rangle$, the point of inflection at the high-temperature side of this maximum may be denoted $T_{m}$ and associated with the crossover between the field-polarized and the paramagnetic regime. \added{For fields above 16~T the point of inflection associated with $T_m$ becomes increasingly shallow. Due to the lack of high-temperature data, $T_m$ is not inferred above 16~T.}

%%%%%%%%%%%%%%%%%%%%%%%%%%%%%%%%%%%%%%%%%%

\subsection{\added[id=CP]{Magnetic Phase Diagrams}}
\label{phasediagram}

\begin{figure}
	\includegraphics[width=1.0\linewidth]{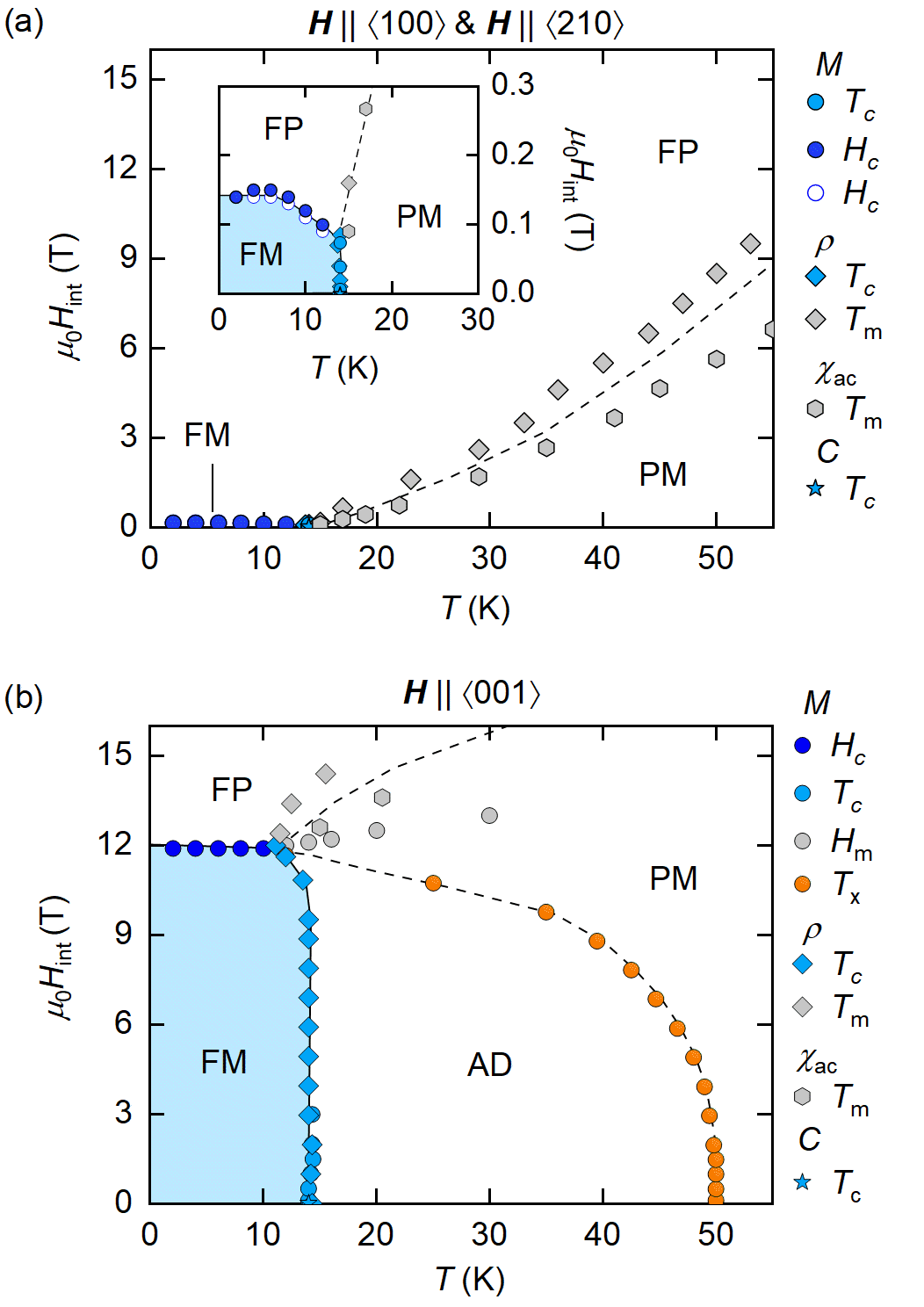}
	\caption{\label{figure11}Magnetic phase diagrams for magnetic fields along the major crystallographic axes. To obtain estimates for the internal field values, demagnetization effect were taken into account. (a)~Diagram for field in the magnetic easy basal plane, combining data measured for field parallel to $\langle100\rangle$ (solid symbols) and $\langle210\rangle$ (open symbols). Ferromagnetic ordered (FM), paramagnetic (PM), and field-polarized (FP) regimes are distinguished. Inset: Diagram for small fields. (b)~Diagram for field along the magnetic hard $\langle001\rangle$ axis. An anisotropy-dominated (AD) regime is observed at temperatures above the long-range ordered state.}
\end{figure}

The magnetic phase diagrams shown in Fig.~\ref{figure11} were inferred from the combined data reported in this paper. Magnetic field values are corrected for demagnetizing fields, allowing to combine data recorded \replaced{on}{in} samples with different geometry, namely the magnetization and resistivity data. Circles, diamonds, hexagons, and stars denote data points inferred from measurements of the magnetization, the electrical resistivity, the ac susceptibility, and the specific heat, respectively. Solid and dashed lines mark phase transitions and crossovers.

For field within the easy magnetic basal plane, shown in Fig.~\ref{figure11}(a), the magnetic phase diagram is that of a soft ferromagnet with a transition temperature $T_{c} = 14$~K.  Results for field parallel to $\langle100\rangle$ (solid symbols) and $\langle210\rangle$ (open symbols) highly resemble each other, consistent with vanishingly small in-plane anisotropy. As the internal field exceeds the very weak in-plane anisotropy, the behavior is characteristic of domain rearrangement processes in a material with multi-domain \replaced[id=CR]{ferromagnetic}{ferromagntic} order (FM) in which a field-polarized regime (FP) is induced. At high temperatures, a paramagnetic regime (PM) is observed.

For field parallel to the hard magnetic $\langle001\rangle$ axis, shown in Fig.~\ref{figure11}(b), the properties are consistent with ferromagnetic order up to high magnetic fields, featuring a spin-flip transition into the field-polarized state at $\mu_{0}H_{c}^{\langle001\rangle} = 12$~T. Above $T_{c}$  a magnetic anisotropy-dominated paramagnetic regime (AD) may be distinguished below ${\sim}50$~K. This regime may be described by a reduced dimensionality of the fluctuation spectrum, where the magnetic moments are essentially confined to the hexagonal basal plane. No additional signatures are observed for magnetic field within the basal plane suggestive of antiferromagnetic correlations or a more complex magnetic state beyond easy-plane ferromagnetism. At higher temperatures, the effects of the magnetic anisotropy are overcome and the magnetic moments may fluctuate in all three dimensions, characteristic of a magnetically isotropic paramagnetic regime. A crossover line, as inferred from a maximum in the magnetization, separates the anisotropy-dominated and the isotropic paramagnetic regime.

At high fields, the magnetic moments are aligned along the field direction, supporting a field-polarized regime, where the crossover towards the paramagnetic state may be tracked in various quantities. In the magnetization, the ac susceptibility, and the electrical resistivity, the crossover temperatures and fields follow the same trends, namely they increase with increasing temperature, but differ in their absolute values. 

Regardless of the precise definition, this crossover as well as the crossover between the anisotropy-dominated and the paramagnetic regime both emanate from the field-induced spin-flip transition at 12~T and 12~K, suggesting the presence of a tricritical point in the magnetic phase diagram of ErB$_{2}$. Furthermore, when considering only magnetic field values above the spin-flip transition at 12~T, the magnetic phase diagram for field along the hard axis resembles strongly that for field along the easy axes.

%These three regimes lacking magnetic long-range order are separated from each other by a broad, triangular-shaped crossover area in which the characteristics of the magnetic fluctuation spectrum smoothly change as a function of temperature or field. Crossover lines inferred from magnetization measurements mark the outer boarders of this crossover regime.

%PM: paramagnetic regime with Heisenberg-like moments fluctuating in three dimensions
%AD: fluctuating regime with XY-like moments confined to two dimensions (basal plane, perpendicular to the field)
%FP: field-polarized regime with Ising-like moments aligned along one dimension (the field direction)
%a triangular crossover regime between three regimes with distinctly different fluctuation characteristics emerging in the presence of strong magnetocrystalline anistropy

%%%%%%%%%%%%%%%%%%%%%%%%%%%%%%%%%%%%%%%%%%

\section{Discussion}
\label{discussion}

\added[id=CP]{The discussion of our results is organized as follows. We begin with a summary of key qualitative findings, defining the basic experimental phenomenology of ErB$_2$. This is followed by basic considerations of the electronic structure providing first insights on the nature of the Er moment. Next, an in-depth discussion of the magnetic properties is presented, arguing that the properties may be quantitatively accounted for by Er$^{3+}$ local moments in the presence of CEFs and van Vleck contributions. The discussion closes with a summary of miscellaneous aspects, ruling out putative scenarios such as mixed valence, strong hybridisation of the Er moments with the conduction bands, modulations of the magnetic order, and zero-point and quantum fluctuations.}

\subsection{\added[id=CP]{Qualitative phenomenology}}

\added[id=CP]{Starting at room temperature, we observe in our study of high-quality single crystal ErB$_2$ a pronounced Curie-Weiss susceptibility in the hexagonal basal plane characteristic of a paramagnetic state down to a well defined thermodynamic phase transition. For decreasing temperature the Curie-Weiss moment exhibits the same effective value at all temperatures down to the phase transition. In comparison, the susceptibility along the crystallographic axis normal to the hexagonal basal plane displays a Curie-Weiss dependence at high temperatures, where the Curie-Weiss moment is comparable to that in the hexagonal plane. Yet, for decreasing temperature a broad maximum emerges well above the magnetic ordering temperature. Thus the paramagnetic state develops a pronounced magnetic anisotropy for decreasing temperature, characteristic of an easy magnetic plane and a hard magnetic axis within and normal to the hexagonal basal plane, respectively.}

\added[id=CP]{As a function of magnetic field and temperature, the magnetization process with respect to zero field and in the crystallographic hexagonal plane is monotonic without hysteresis and additional fine structure. Under magnetic fields up to 30\,T the magnetization increases weakly without fine structure such as changes of slope or curvature. \replaced[id=CR]{In comparison, for the axis normal to the hexagonal basal plane, the magnetization remains very small up to a pronounced transition around 12\,T, above which it flattens with no further features up to 30\,T, the highest field studied.}{In comparison, for the axis normal to the hexagonal basal plane, the magnetization is initially very low up to a pronounced transition around 12\,T followed by a featureless increase up to 30\,T, the highest fields studied.} As a function of increasing and decreasing fields, the pronounced transition does not exhibit any fine-structure or hysteresis.}

\added[id=CP]{The specific heat displays the usual lattice contributions at high temperatures and a pronounced Schottky anomaly with decreasing temperature followed by a single, well-defined second order phase transition consistent with the susceptibility and magnetization. The electrical resistivity is characteristic of a high-purity metallic state, featuring also the characteristics of the magnetic phase transition at low temperatures consistent with the thermodynamic properties.}

\added[id=CP]{Taken together, the susceptibility, magnetization, specific heat, and electrical resistivity of ErB$_{2}$ for fields applied along the major crystallographic axes are phenomenologically characteristic of a paramagnetic state with a strong easy-plane magnetic anisotropy that emerges at low temperatures. The basal plane is magnetically easy with vanishingly small in-plane anisotropy and $\langle001\rangle$ is the hard magnetic axis. Within this highly anisotropic paramagnetic state long-range easy-plane ferromagnetic order develops in a single well-defined phase transition.}

%%%%%%%%%%%%%%%%%%%%%
\subsection{\added[id=CP]{Electronic structure}}
 
\replaced[id=CR]{As the starting point for the discussion of the electronic structure we assume an Er$^{3+}$ ionic configuration, noting that the trivalent state of Er is energetically well separated from both the divalent Er$^{2+}$ (4$f^{12}$) and the tetravalent Er$^{4+}$ (4$f^{10}$) configuration. Formally, the Er$^{3+}$ thereby donates the two $6s$ and one $4f$ electron of the atomic configuration $\rm {[Xe]} 4f^{12} 6s^{2}$.~\cite{Johansson_1979, Strange_1999} In contrast to MgB$_2$, where holes in the B $\sigma$ bands are present at the Fermi level~\cite{Boeri_2002}, the $\sigma$ bands of diborides with trivalent metals are typically completely or almost completely filled, and metal $d$ states contribute strongly at $E_F$~\cite{Medvedeva_2001,Nakamura_2002}. For ErB$_2$, first-principles calculations accordingly find Er-$5d$ states hybridized with B $2p$ at the Fermi level, while the $4f$ states remain localized well below $E_F$~\cite{CasianoJimenez_2014}, as also reported for TmB$_2$~\cite{2009_Mori_PhysRevB}. The boron layer thus forms a covalently bonded, graphene-like honeycomb network (B--B distance $\approx 1.9\,{\rm \AA}$~\cite{2022_Bauer_PhysStatusSolidiB}) into which the valence electrons of the Er are released.}{\added[id=CP]{As the starting point for the discussion of the electronic structure we assume an Er$^{3+}$ ionic configuration, noting that Er lacks any shell stabilization for Er$^{2+}$ (4$f^{12}$) or Er$^{4+}$ (4$f^{10}$), i.e., the promotion energies are too high for a mixed valence state. Formally, the Er$^{3+}$ thereby donates the two $6s$ and one $4f$ electron of the atomic configuration $\rm {[Xe]} 4f^{12} 6s^{2}$.~\cite{Johansson_1979, Strange_1999} Consistent with the typical situation in the AlB$_2$/C32 diborides the B does not support a well-defined ionic charge in the strict sense. Rather, the boron atoms form a covalently bonded honeycomb layer (graphene-like, B--B distance $\approx 1.9\,{\rm \AA}$~\cite{2022_Bauer_PhysStatusSolidiB}) into which the valence electrons of the Er are released. \replaced[id=CR]{}{In analogy with MgB$_2$ the B$^{2-}$ honeycomb $\pi^*$ band is exactly half filled, which gives rise to the superconducting band structure.\added[id=CP]{The additional electron per formula unit in ErB$_2$ goes predominantly into Er-$5d/6s$-derived conduction bands and not in the boron $\pi^*$ system, while the covalent $\sigma$ and $\pi$ bands of the boron layer look largely as they do in MgB$_2$. This is supported by a Bader/Mulliken analysis from DFT, which typically give partial charges of about -0.3 to -0.7 e for boron in these structures, thus clearly more electronegative than the metal, but far from the formal ionic charge.}}}}

\added[id=CP]{Consistent with this picture, the low residual resistivity of a few ${\rm \mu\Omega\,cm}$ and the small Sommerfeld coefficient of the order a few $\rm{mJ\cdot mol^{-1}\cdot\,K^{-2}}$ correspond to a good metallic conduction band with weak correlations, a localized 4f$^{11}$ moment on the Er site, and a covalently bonded B layer. This is supported further by the well-defined Schottky anomaly, reflecting the CEF splitting of the $^4I_{15/2}$ multiplet of Er$^{3+}$ as a well-defined local ion with the full Hund's-rule configuration.}

%%%%%%%%%%%%%%%%
\subsection{\added[id=CP]{Magnetic properties}}

\added[id=CP]{The magnetic properties of ErB$_2$ may be well accounted for by Er$^{3+}$ local moments of a $4f^{11}$ configuration, consistent with the considerations of the electronic structure. In the Hund's-rule ground state $^4I_{15/2}$ ($J = 15/2$, $L = 6$, $S = 3/2$, $g_J = 6/5$) the free Er$^{3+}$ ion has an effective paramagnetic moment $\mu_{\rm eff} = g_J \cdot \sqrt{J(J+1)} \approx 9.58\,\mu_{\rm B}$ and a saturation moment $\mu_{\rm sat} = g_J \cdot J = 9.0\,\mu_{\rm B}$. In comparison, the quantitative value of $\mu_{\rm eff}$ between 9.13 and $9.64\,\rm{\mu_B}\,f.u.^{-1}$ inferred from the Curie--Weiss dependence corresponds practically to the Hund's-rule free-ion value. \replaced[id=CR]{Further, the magnetization at high field, $m_\mathrm{hf}^{\langle100\rangle} \approx 7.8~\mu_{\mathrm{B}}\,\mathrm{f.u.}^{-1}$ within the basal plane and $m_\mathrm{hf}^{\langle001\rangle} \approx 8.3~\mu_{\mathrm{B}}\,\mathrm{f.u.}^{-1}$ along the hard $\langle001\rangle$ direction at 30~T (cf.\ Fig.~\ref{figure5}), is reduced relative to the free-ion value in both cases. Within the basal plane the magnetization is still unsaturated at 30~T.}{Further, the magnetization at high field $m_\mathrm{hf}^{\langle100\rangle} \approx 7.8~\mu_{\mathrm{B}}\,\mathrm{f.u.}^{-1}$ within the basal plane and $m_\mathrm{hf}^{\langle001\rangle} \approx 8.3~\mu_{\mathrm{B}}\,\mathrm{f.u.}^{-1}$ along the hard $\langle001\rangle$ direction at 30~T (cf.\ Fig.~\ref{figure5}) are both reduced relative to the free-ion value and unsaturated.} Remarkably, the moment at high field for the hard axis above the spin-flip transition {\it exceeds} the moment for the easy plane. Finally, the magnetic entropy tends towards ${\rm 4R\cdot \ln 2 = R\ln 16}$ in excellent agreement with the 16-fold degeneracy of the $^4I_{15/2}$ multiplet ($J = 15/2$).}

\added[id=CP]{In a crystalline environment such as ErB$_2$, the CEF lifts the 16-fold degeneracy of the $^4I_{15/2}$ multiplet of the Er$^{3+}$. At low temperatures only the lowest Kramers doublet is occupied, and its wavefunction is in general a linear combination of $\vert M_J \rangle$ states with $\vert M_J\vert \leq 15/2$. The expectation value $\langle J_x\rangle$ in the CEF ground state is therefore smaller (or, in the limiting case, even considerably smaller) than $J = 15/2$. This directly implies a reduction of the saturation moment. Even in simple tetragonal or hexagonal environments, moderate CEF parameters suffice to produce values around $7 - 8.5\,\mu_{\rm B}$. \replaced[id=CR]{Additional intermediate-coupling corrections, i.e., an admixture of other terms with $J = 15/2$ (mainly $^{2}K_{15/2}$, of the order of a few percent~\cite{Chai2016}) into the $LS$-pure $^{4}I_{15/2}$ ground multiplet, reduce $g_{J}$ from the $LS$ value $6/5$ by only a few tenths of a percent and are negligible at the accuracy of the present data.}{Additional intermediate-coupling corrections, i.e., an admixture of the $^{4}I_{13/2}$ multiplet into the $LS$-pure $^{4}I_{15/2}$ ground state, reduce $g_{J}$ from the $LS$ value $6/5$ on the level of percent and act here as a fine correction.}}

\replaced[id=CR]{The experimental observation that the magnetization at high fields within the easy plane is unsaturated, while the magnetization along the magnetically hard $\langle001\rangle$ direction \emph{exceeds} the in-plane value, is consistent with the CEF-plus-Van-Vleck picture outlined here. Consider the simplest scenario of an easy-plane CEF ground-state doublet, characterized by a large moment within the basal plane and a vanishing moment along $\langle001\rangle$, and an excited Kramers doublet with a large moment along $\langle001\rangle$. For a magnetic field applied along $\langle001\rangle$, the ground-state doublet experiences essentially no linear Zeeman shift, whereas the excited doublet splits, with its lower Zeeman branch shifting down in energy in proportion to its large $c$-axis moment. At $\mu_{0}H_{c}$ this branch crosses the ground-state doublet, and the magnetization along $\langle001\rangle$ is subsequently set by the moment of the excited level rather than by the ground-state doublet. This accounts naturally for the pronounced transition at $\mu_{0}H_{c} \approx 12$~T and for the large moment along the hard axis at high fields. A simple estimate supports this scenario quantitatively. The Zeeman energy of a level with moment $m_\mathrm{hf}^{\langle001\rangle} \approx 8.3~\mu_{\mathrm{B}}\,\mathrm{f.u.}^{-1}$ at $\mu_{0}H_{c} = 12$~T amounts to $|E^\mathrm{Zee}| = m_\mathrm{hf}^{\langle001\rangle}\,\mu_{0}H_{c} \approx 5.8$~meV, in good agreement with the splitting of $\approx 6$~meV inferred from the Schottky anomaly. In contrast, for fields within the easy plane no such level crossing occurs, and the finite slope of the magnetization up to 30~T reflects Van Vleck contributions from the field-induced admixture of excited CEF levels. A determination of the CEF level scheme, and hence a quantitative disentangling of the individual contributions, requires direct spectroscopic evidence e.g., by inelastic neutron scattering, which is beyond the scope of the present study.}{\added[id=CP]{The experimental observation that the magnetization at high fields is unsaturated and along the magnetically hard $\langle001\rangle$ direction \emph{exceeds} the in-plane value is consistent with the CEF-plus-Van-Vleck picture outlined here. Above the spin-flip transition at $\mu_{0}H_{c} = 12$~T the magnetization along $\langle001\rangle$ is no longer set by the easy-plane $g_{\perp}$ of the CEF ground-state doublet alone, but receives a sizeable Van Vleck contribution from admixed excited CEF levels, which is largest along the hard axis where the CEF--Zeeman competition is strongest. The persistent finite slope of the in-plane magnetization up to 30~T reflects a similar, weaker Van Vleck contribution that has not yet saturated in the available field range. Determination of the exact CEF level scheme and thus disentangling the individual contributions on a quantitative level would require direct spectroscopic evidence e.g., by inelastic neutron scattering, which is beyond the scope of the present study.}}

\added[id=CP]{Thus, the comparatively small differences of $\mu_{\rm eff}$ and $m_\mathrm{hf}$ relative to the free ion value together with the lack of saturation, strong magnetic anisotropy, and the Schottky anomaly in the specific heat may be explained well by CEF plus Van Vleck contributions. In contrast, as discussed in further detail below and based on the phenomenology observed experimentally, the presence of modulations of the magnetic structure, dominant zero-point and quantum fluctuations, or Er mixed-valence appears highly unlikely.}

%A more comprehensive understanding of the crystal field level scheme and its impact on the magnetic behavior would, however, require direct spectroscopic measurements, such as inelastic neutron scattering—which lie beyond the scope of the present study.

%%%%%%%%%%%%%%%%%%%%%
\subsection{\added[id=CP]{Further aspects}}

\added[id=CP]{In the following we address further well-established mechanisms resulting in reduced values of the Curie-Weiss moment, $\mu_{\rm eff}$, and the high-field magnetization, $m_\mathrm{hf}$, relative to the free ion value. Namely, the formation of mixed valence, hybridization of the magnetic moments with the conduction electrons, modulated forms of magnetic order, and zero-point and quantum fluctuations. While none of these mechanisms appears to be relevant in ErB$_2$, it is instructive to summarize the precise arguments supporting this conjecture.}

\added[id=CP]{Starting with mixed-valence phenomena, these typically concentrate on those elements for which one of the two valence states involved possesses a particularly stable 4f configuration. Key examples include $4f^0$ (Ce$^{4+}$), $4f^7$ (Eu$^{2+}$, Tb$^{4+}$), $4f^{14}$ (Yb$^{2+}$) and, to a lesser extent, configurations close to empty, half or full filling such as Sm$^{2+}$ ($4f^6$), Tm$^{2+}$ ($4f^{13}$) and Pr$^{4+}$ ($4f^1$). For Er, the trivalent configuration $4f^{11}$ is far removed from any of these special situations. A hypothetical Er$^{2+}$ species would have $4f^{12}$, an Er$^{4+}$ species $4f^{10}$, both configurations without any appreciable Hund's-rule bonus. 
\replaced[id=CR]{Correspondingly, the energy difference between the divalent and the trivalent state amounts to about 1.6\,eV, twice the value for Tm (0.8\,eV). Similarly, converting a trivalent Er site into a tetravalent one costs about 5\,eV~\cite{Johansson_1979}. X-ray photoemission places the occupied $4f$ states of Er metal (final state $4f^{10}$) about 4.7 eV below $E_F$, and bremsstrahlung isochromat spectroscopy the unoccupied $4f$ states (final state $4f^{12}$) about 2.2 eV above $E_F$~\cite{Johansson_1979}.}{Correspondingly, the promotion energies for Er$^{2+}$ (4$f^{12}$) or Er$^{4+}$ (4$f^{10}$) are high. The difference between the divalent and the trivalent promotion energy is about \replaced[id=CR]{0.5-1.4\,eV}{1.5-2\,eV} larger for Er than for Tm, and \replaced[id=CR]{2.1-2.7\,eV}{2-3\,eV} larger than for Yb. A comparably large difference exists between the trivalent and the tetravalent state.Consequently, according to photoemission, the occupied $4f^{11}$ multiplet of Er intermetallics lies roughly \replaced[id=CR]{5\,eV}{5-10\,meV} below the Fermi energy, and the f-hole level, which would correspond to a transition to Er$^{4+}$, \replaced[id=CR]{roughly 2\,eV}{several eV} above $E_F$.~\cite{Johansson_1979}}}  

\added[id=CP]{While mixed valence practically does not occur in Er compounds, several exceptions are conceivable in principle that have not been observed, however. (i) Extremely high pressures may shift valence promotion energies by a few eV, where pressure-induced valence transitions in Er have so far not been reported. (ii) In molecular complexes with very strong donor ligands (e.g. $Cp'$-based compounds~\cite{MacDonald2012}) formally Er$^{2+}$ species can be generated, but these are isolated chemical cases and not intermetallic mixed-valence states. (iii) In strongly strained thin films or at interfaces, local symmetry reduction may in principle influence excitation energies, without this having so far been demonstrated for Er.}

%%%%%%%
\replaced[id=CR]{As discussed in Sec.~IV~B, the $4f$ states of Er do not hybridize appreciably with the conduction bands, such that a reduction of the moment by $4f$ hybridization is highly unlikely. The remaining coupling, the intra-atomic $4f$--$5d$ exchange, polarizes the $5d$ conduction states parallel to the $4f$ spin, which for $J \parallel S$ slightly enhances rather than reduces the total moment~\cite{Roeland_1975,Brooks_1991}. Such contributions of a few tenths of $\mu_\mathrm{B}$ would increase rather than reduce the high-field moment of ErB$_2$ and hence cannot account for the observed reduction.}{\added[id=CP]{Hybridization of the 4$f$ shell with conduction electrons (4$f$--5$d$ mixing) may reduce the moment by a few percent in Er compounds with substantial $d$-character at the Fermi level, such as Er--Co, Er--Ni, or Er--Ru intermetallics.}}
 
\replaced[id=CR]{Similarly, the large separation of the $4f^{11}$ configuration from both the divalent and the tetravalent configuration~\cite{Johansson_1979}, together with the localized nature of the 4$f$ states, renders a Kondo-like reduction of the moment highly unlikely. The antiferromagnetic exchange generated by virtual charge fluctuations between the 4$f$ shell and the conduction electrons, $J_K \propto -|V|^2/\Delta\epsilon$~\cite{Schrieffer_1966,Coqblin_1969}, is negligible for excitation energies $\Delta\epsilon$ of several eV, such that the coupling to the conduction electrons is dominated by the ferromagnetic intra-atomic exchange discussed above, which also gives rise to the RKKY interaction. Consistent with this, Kondo-like behavior in 4$f$ systems has been observed for compounds of Ce, Pr, Sm, Eu, Tm, and Yb~\cite{Maple_1978}, whereas to our knowledge no Kondo effect has been reported for Er intermetallics. This is also consistent with the low residual resistivity and the small Sommerfeld coefficient of ErB$_2$.}{\added[id=CP]{\replaced[id=CR]{In ErB$_2$, the 4$f$ states are far removed from the Fermi level, with excitation energies of several eV towards both the Er$^{2+}$-like and the Er$^{4+}$-like configuration~\cite{Johansson_1979}. Band-structure calculations for the isostructural TmB$_2$ show that the states at $E_F$ are of mixed B-2p and rare-earth 5d character, whereas the 4$f$ states are strongly localized and do not mix noticeably with the conduction bands, such that the 4f moments couple via the RKKY interaction~\cite{2009_Mori_PhysRevB}. Consequently, the Kondo coupling is expected to be weak and a Kondo-like reduction of the moment is negligible, consistent with the low residual resistivity, and the small Sommerfeld coefficient.}{In ErB$_{2}$ the conduction states are dominated by Er-derived 5$d$/6$s$ and B 2$p$ contributions, such that a Kondo-like reduction is expected to play a minor role consistent with the low residual resistivity and the small Sommerfeld coefficient.} Framed in the periodic Anderson model of a mixed-valence regime, the $f$ level (or the $f$-hole level) would have to approach $E_F$ to within a few $T_K,$ and the hybridization $V$ with the conduction band would have to compete with the promotion energy. For Er, $\Delta \epsilon \gg V$ and the system remains deep in the regime of stable integer valence. The Kondo effect is in principle possible in Er intermetallics, but with Kondo temperatures usually well below 1\,K it is so weak that the magnetic phenomenology is dominated by the crystal field and the exchange interaction.} \added[id=CR]{\textit{Can not find evidence for the below 1K claim. In fact I cannot find any Kondo effect reported for Er based compounds}}}

\added[id=CP]{Further, many Er intermetallics order magnetically in modulated structures, notably sinusoidally modulated, helical, multi-$q$ or incommensurate forms of order~\cite{Jensen_1991,Lynn_1997,Fruhling_2024,Garnier_2000}. In the presence of such modulations, the uniform magnetization is reduced. However, as a function of magnetic field, the behavior around zero magnetic field typically exhibits additional features beyond the characteristics of a simple ferromagnetic hysteresis loop. Moreover, under magnetic field such modulated structures display field-induced transitions such as spin-flop transitions (frequently involving several steps). No such features are observed in our measurements, providing strong empirical evidence against modulations or superstructures of an otherwise ferromagnetic state.}

\added[id=CP]{Finally, quantum and zero-point fluctuations reducing the ordered moment, which can be sizeable in geometrically frustrated or low-dimensional Er systems, are expected to be small in ErB$_2$. In three-dimensional RKKY-coupled magnets the effect is generically modest, and the small frustration parameter $f_{\langle001\rangle} = 2.1$ extracted in ErB$_2$ (cf. Sec.~\ref{susceptibility}) indicates that competing interactions do not significantly suppress the ordering.}

%%%%%%%%%%%%%%%%%%%%%%%%%%%%%%%%%%%%%%%%%%

\section{Conclusions}
\label{conclusion}

In summary, we reported an investigation of the low-temperature properties of \added[id=CP]{high-quality} single crystals of the hexagonal C32 diboride ErB$_{2}$. Based on comprehensive measurements of the ac susceptibility, magnetization, specific heat, and electrical resistivity, we find 
\replaced[id=CP]{a paramagnetic state at room temperature with a strong easy-plane magnetic anisotropy in the crystallographic hexagonal basal plane that emerges below $\sim50\,{\rm K}$. At a transition temperature of $T_{c} = 14$~K pronounced easy-plane ferromagnetism with vanishingly small in-plane magnetic anisotropy stabilizes in a weakly correlated metallic state. Both the qualitative phenomenology as well as the quantitative characteristics up to high magnetic fields of 30\,T are in excellent agreement with Er$^{3+}$ local moments in the presence of crystal electric fields and \replaced[id=CR]{}{weak intermediate-coupling} Van Vleck contributions. Taken together, our study identifies ErB$_2$ as an example par excellence featuring the simplest form of long-range magnetic order: ferromagnetism.}
{easy-plane ferromagnetic order below $T_{c} = 14$~K. Curie--Weiss fits to the ac susceptibility indicate ferromagnetic coupling along $\langle100\rangle$ and $\langle210\rangle$, antiferromagnetic coupling along $\langle001\rangle$, and fluctuating moments that are consistent with the free-ion value of Er$^{3+}$.}

\deleted{However, the reduced high-field magnetization relative to the free-ion value, the absence of saturation in the in-plane magnetization, and the pronounced Schottky anomaly in the specific heat collectively suggest that the Hund's rule ground state is split by the CEF. However, a lack of saturation in the magnetization at high fields, typically attributed to itinerant magnetism, suggests that the 4f electrons may be hybridized with d electrons at the Fermi level.}

\deleted[id=CP]{A strong magnetic anisotropy dominates the magnetic properties at low temperatures, where the hexagonal basal plane is magnetically easy with a vanishingly small in-plane anisotropy and a magnetic hard $\langle001\rangle$ axis. For field within the basal plane, the magnetic properties are reminiscent of a soft ferromagnet that is polarized in fields above a few tenth of a Tesla, while for field along $\langle001\rangle$ a spin-flip transition into the field-polarized state is observed at ${\sim}12$~T. The magnetic anisotropy dominates the paramagnetic state up to $T_{x} = 50$~K, where magnetic fluctuations are essentially confined to the basal plane. 
%Under increasing field or temperature, crossovers into the field-polarized and the paramagnetic regime are induced, giving rise to a triangular-shaped crossover area between these three regimes of distinctly different fluctuation characteristics.% Future studies ... These findings identify ErB$_{2}$ as a model system for studying the interplay of ... in the presence of strong magnetocrystalline anisotropy
As a function of magnetic field and temperature, crossover between the field-polarized, the paramagnetic, and the anisotropy-dominated regime may be distinguished featuring distinct changes in the spectrum of magnetic fluctuations.}

\begin{acknowledgments}
We wish to thank A.\ Engelhardt, S.\ Mayr, W.\ Simeth, and M.\ Stekiel for fruitful discussions and assistance with the experiments. \added{This work was supported by HFML-RU/NWO-I, member of the European Magnetic Field Laboratory (EMFL).} This study was funded by the Deutsche Forschungsgemeinschaft (DFG, German Research Foundation) under TRR80 (From Electronic Correlations to Functionality, Project No.\ 107745057), TRR360 (Constrained Quantum Matter, Project No. 492547816), SPP2137 (Skyrmionics, Project No.\ 403191981, Grant PF393/19), and the excellence cluster MCQST under Germany's Excellence Strategy EXC-2111 (Project No.\ 390814868). Financial support by the European Research Council (ERC) through Advanced Grants No.\ 291079 (TOPFIT) and No.\ 788031 (ExQuiSid) is gratefully acknowledged.
\end{acknowledgments}

\bibliography{ExportedItems}

\end{document}